\documentclass{PoS}

\usepackage{amssymb,amsmath,amsfonts,amsthm,bbold}
\usepackage{mathrsfs,mathtools}
\usepackage{graphicx}
\usepackage{empheq} 
\usepackage{verbatim}
\usepackage{stmaryrd}
\usepackage{textcomp}
\usepackage{minibox} 
\usepackage{url}

\DeclareMathOperator{\extdm}{d}
\newcommand{\extd}{\extdm \!}

\newcommand{\unity}{1\hspace{-0.243em}\text{l}}
\newcommand{\p}{\partial}

\newtheorem{theorem}{Theorem}[section]

\providecommand{\Lt}{{\tt L}}
\renewcommand{\Lt}{{\tt L}}
\providecommand{\Mt}{{\tt M}}
\renewcommand{\Mt}{{\tt M}}

\title{Canonical Charges in Flatland}

\ShortTitle{An Introduction to Asymptotic Symmetry Analysis}

\author{\speaker{Max Riegler\footnote{Speaker.} and C{\'e}line Zwikel}\\
        Universit{\'e} libre de Bruxelles\\
        E-mail: \email{max.riegler@ulb.ac.be}, \email{czwikel@ulb.ac.be}}

\abstract{In this series of lectures we give an introduction to the concept of asymptotic symmetry analysis with a focus on asymptotically flat spacetimes in 2+1 dimensions. We explain general ideas of quantizing gauge theories and then apply these ideas to gravity both in the metric as well as the Chern-Simons formulations. This enables one to compute the asymptotic symmetries of given gravitational configurations that in turn act as the basic underlying symmetries of a possible dual quantum field theory in the context of holography. We also briefly elaborate on the concept of "soft hair" excitations of black holes in this context.}

\FullConference{XIII Modave Summer School in Mathematical Physics\\
                10--16 September 2017\\
                Modave, Belgium}

\begin{document}

\section*{Note to the Reader}

These lecture notes are intended for a six hours lecture course given at the XIII Modave Summer School in Mathematical Physics. The main purpose of the lecture notes is to give graduate students the possibility to obtain a basic working knowledge of constrained Hamiltonian systems, the importance of canonical boundary charges and asymptotic symmetries in general relativity. Furthermore we want to provide a foundation for understanding the basic ideas underlying the concept of black holes with soft hair and the relation to the black hole information paradox. These lecture notes are kept short and simple on purpose and focus on the main ideas and concepts underlying the topics that are discussed. For a further in depth study of all these interesting topics we point out literature for further reading in the main body of the lecture notes.

\section{Introduction and Motivation}

The concept of asymptotic symmetries plays a prominent role in general relativity. In short, these asymptotic symmetries of a given spacetime are given by the gauge transformations that preserve the asymptotic structure of a given set of boundary conditions. The existence of such asymptotic symmetries usually also means that there are corresponding conserved charges within the bulk of the spacetime under consideration. As such these symmetries allow one to define for example notions of e.g. mass or angular momentum for observers at asymptotic infinity. Furthermore, these symmetries also play a prominent role for the underlying symmetries of a putative dual quantum field theory in the context of the holographic principle\footnote{The holographic principle is a conjectured duality between a theory of (quantum)gravity in d+1 dimensions and a quantum field theory in d dimensions that is defined on the boundary of the gravity theory. Maybe the most famous realization thereof is the so called AdS/CFT correspondence \cite{Maldacena:1997re} that involves spacetimes with constant negative curvature i.e. Anti-de Sitter spacetimes and conformal field theories.}.\\
There is an intimate interplay between the existence of global charges and asymptotic symmetries \cite{Abbott:1981ff,Abbott:1982jh} and as such a precise treatment and definition of global charges is of vital importance. This is usually done via a procedure called a canonical analysis that basically treats e.g. gravity as a constrained Hamiltonian system and gives a way of quantizing such a constrained system in the presence of gauge symmetries. This procedure enables one to define the aforementioned canonical charges, their associated Dirac brackets and ultimately also the algebra of asymptotic symmetries if the analysis is performed at the asymptotic boundary of a given spacetime.\\
Another important aspect of the existence or non-existence of canonical charges associated to a given asymptotic symmetry transformation is that they allow one to distinguish \emph{proper} from \emph{improper} gauge transformations.

\paragraph{Proper gauge transformations:}

These are proper gauge transformations in the sense that they do not change the physical state of the system and are in general associated to gauge transformations that have a vanishing canonical charge at the boundary.

\paragraph{Improper gauge transformations:}

These are improper gauge transformations and thus not really gauge transformations in the classical sense as these transformations do change the physical state of the system. These kind of transformations usually have a non-vanishing charge at the boundary.\\
It is exactly the second type of gauge transformations that is of interest for most holographic applications. If a symmetry transformation changes the physical state of a system then this also means that one can use these symmetries to build modules containing various different states in the dual quantum field theory. This is done by starting from a ground state which is usually some kind of vacuum by repeated application of canonical generators of asymptotic symmetries. This allows one to check partially\footnote{Partially in this case means that one can perform checks on the level of the basic symmetries of a putative dual quantum field theory. More intricate checks that go beyond that scope require more detailed knowledge of a possible quantum field theory dual.} the validity of possible holographic conjectures for many cases (especially in three dimensions).\\
This process of performing a canonical analysis is not purely restricted to asymptotic symmetries alone. One can employ the same logic and techniques also for slightly different setups such as e.g. the near-horizon symmetries of black holes \cite{Afshar:2016wfy,Donnay:2015abr}.\\
The importance of asymptotic symmetries with a special focus on the Bondi-Metzner-Sachs ($\mathfrak{bms}$) algebra \cite{Bondi:1962px,Sachs:1962zza} was highlighted recently in the context of the black hole information paradox\footnote{This paradox basically revolves around the question what happens with information that was previously thrown into a black hole during the evaporation process of the black hole.} as well. The authors of \cite{Hawking:2016msc}, Hawking-Perry-Strominger argued that the Hawking radiation that is emitted by a black hole can be decomposed into \emph{hard} and \emph{soft} quanta and that these additional soft quanta that are basically zero energy excitations might provide a new way of tackling the information paradox.\\
The purpose of these lecture notes is to get a basic understanding of asymptotic symmetries, the canonical charges associated to these symmetries and the physical consequences of having both canonical charges and asymptotic symmetries. We will put a focus on three-dimensional gravity for illustrational purposes as well as asymptotically flat spacetimes. The reason for focusing on asymptotically flat spacetimes is that the $\mathfrak{bms}$ algebra appears as the asymptotic symmetry algebra of asymptotically flat spacetimes. This allows one to gain a bit of intuition on the $\mathfrak{bms}$ algebra which will be one of the basic ingredients to get an elemental understanding of the proposal of Hawking-Perry-Strominger.

\section{Preliminaries}

This part\footnote{Please note that his preliminary section is based on the introduction of \cite{Riegler:2016hah}.} of the lecture note is mainly devoted to the basic preliminary knowledge needed in order to follow the main body of the lecture notes. Since the focus of these notes is on gravity in three dimensions we will first review the special properties of gravity theories in three dimensions and will then proceed in describing how to formulate gravity as a Chern-Simons gauge theory.

\subsection{Gravity in 3D}
%
General Relativity in three dimensions is very special in many regards and there are a lot of reasons why it is beneficial to study gravity in this setup, especially if one is interested in general features of holography.\\
First and foremost, gravity in three dimensions is technically much simpler than in four or higher dimensions. For example the Riemann tensor $R_{abcd}$ can be expressed in terms of the Ricci tensor $R_{ab}$, the Ricci scalar $R$ and the metric $\kappa_{ab}$ as
	\begin{equation}
		R_{abcd}=g_{ac}R_{bd}+g_{bd}R_{ac}-g_{ad}R_{bc}-g_{bc}R_{ad}-\frac{1}{2}R(g_{ac}g_{bd}-g_{ad}g_{bc}).
	\end{equation}
Now take into account Einstein's equations
	\begin{equation}\label{eq:EinsteinEquationsSecondOrder}
		R_{\mu\nu}+\left(\Lambda -\frac{R}{2}\right)g_{\mu\nu}=8\pi G T_{\mu\nu},
	\end{equation}
where $G$ is Newton's constant in three dimensions, $\Lambda$ is the cosmological constant and $T_{\mu\nu}$ the energy-momentum tensor which encodes the local energy-momentum distribution. This implies that the curvature of spacetime in three dimensions is completely determined in terms of the local energy-momentum distribution and the value of the cosmological constant. Thus, if there are no matter sources the curvature of spacetime is completely determined by the value of the cosmological constant. This in turn also means that there are no local propagating (bulk-) degrees of freedom i.e. massless gravitons\footnote{This is true for Einstein-Hilbert gravity in three dimensions. One could, however, also consider other gravity theories in three dimensions which allow for (typically massive) gravitons.}.\\
At first sight this sounds like bad news since a theory with no local propagating degrees of freedom seems to be trivial. Luckily, both local and global effects play an important role in (three-dimensional) gravity so that the theory is physically non-trivial. It is also noteworthy that Einstein gravity in three dimensions is a topological theory.\\
Probably the most famous example illustrating this feature is the BTZ black hole solution found by Ba\~nados, Teitelboim and Zanelli \cite{Banados:1992wn,Banados:1992gq}. This black hole solution is locally AdS, but at the boundary of the AdS spacetime it is characterized by canonical charges\footnote{We will see later on in these lecture notes what these canonical charges are precisely.} that differ from the usual AdS vacuum. In addition the BTZ black hole has a horizon, singularity and exhibits an ergoregion in general.\\
In \cite{Brown:1986nw} Brown and Henneaux presented boundary conditions for three dimensional gravity, whose corresponding canonical charges generate two copies of the Virasoro algebra. This ultimately lead to the (holographic) conjecture that AdS in three dimensions can equivalently be described by a two-dimensional conformal field theory located at the boundary of AdS \cite{Witten:2007kt}.\\
Since gravity in three dimensions is a purely topological theory one might expect that this theory can also be formulated in a way that makes its topological character explicit e.g. a Chern-Simons formulation. We will review Chern-Simons formulations and its properties in Section~\ref{ch:ChernSimonsFormulationofGravity}. Before doing so it will be instructive to explain how one has to formulate gravity in three dimensions in order to be able to rewrite the Einstein-Hilbert action
	\begin{equation}\label{eq:EHSecondOrderAction}
		I_{\textnormal{EH}}=\frac{1}{16\pi G}\int_{\mathcal{M}}\extd^3x\sqrt{-g}\left(R-2\Lambda\right),
	\end{equation}
where $g\equiv\det g_{\mu\nu}$, as a Chern-Simons action\footnote{There is an excellent set of lecture notes going a little bit into more detail on this topic by Laura Donnay \cite{Donnay:2016iyk}.}.\\
The action \eqref{eq:EHSecondOrderAction} takes as the fundamental dynamic field the symmetric tensor $g_{\mu\nu}$ which acts as a symmetric bilinear form on the tangent space of the manifold $\mathcal{M}$. Writing the metric in a given basis thus does not necessarily mean that this basis is orthonormal at each given point of spacetime. For many purposes it is, however, advantageous to have a notion of a local orthonormal laboratory frame i.e. a family of ideal observers embedded in a given spacetime. Such a family of ideal observers can be introduced in general relativity via frame fields $e^a=e^a{}_\mu\extd x^\mu$, which are often also called \emph{vielbein}. This frame field is a function of the spacetime coordinates $x^\mu$ and carries spacetime indices, which will be denoted by Greek letters $\mu,\nu,\ldots$ and internal local Lorentz indices denoted by Latin letters $a,b,\ldots$. The frame fields $e^a$ and the metric $g_{\mu\nu}$ are related by
\begin{equation}\label{Intro:MetricDreibeinRel}
		g_{\mu\nu}=e^a{}_\mu e^b{}_\nu\eta_{ab},
	\end{equation}
where $\eta_{ab}$ is the 2+1 dimensional Minkowski metric with signature ($-,+,+$). In this formulation local Lorentz indices can be raised and lowered using the Minkowski metric $\eta_{ab}$, while spacetime indices are raised and lowered using the spacetime metric $g_{\mu\nu}$.\\
The big advantage of using a formulation in terms of frame fields is that one now can very easily promote objects from a flat, Lorentz invariant setting to a description in a coordinate invariant and curved background\footnote{One example would be a formulation of the Dirac equation in curved backgrounds.}. Take for example some object $V^a$ which transforms under local Lorentz transformations $\Lambda\left(x^\mu\right)^a{}_b$ like the components of a vector,
	\begin{equation}
		\tilde{V}^a=\Lambda\left(x^\mu\right)^a{}_bV^b.
	\end{equation}
Then one can easily describe this object in a curved background using the frame field\footnote{To be more precise this is the inverse of the frame field $e^a{}_\mu$ defined by $e^a{}_\mu e_a{}^\nu=\delta_\mu^\nu$.} as
	\begin{equation}
		V^\mu=e_a{}^\mu V^a.
	\end{equation}
Local Lorentz invariance of the frame fields also means that there should be a gauge field associated to that local Lorentz invariance. This gauge field is the spin connection $\omega^{ab}=\omega^{ab}{}_\mu \extd x^\mu$ with $\omega^{ab}{}_\mu=-\omega^{ba}{}_\mu$ which also allows one to define a covariant derivative acting on generalized tensors i.e. tensors which have both spacetime and Lorentz indices as
	\begin{equation}
		\mathcal{D}_\mu V^a{}_\nu=\partial_\mu V^a{}_\nu+\omega^a{}_{b\mu}V^b{}_\nu-\Gamma^\sigma{}_{\nu\mu}V^a{}_\sigma,
	\end{equation}
where $\Gamma^\sigma{}_{\nu\mu}$ denotes the affine connection associated to the metric $g_{\mu\nu}$
	\begin{equation}
		\Gamma^\sigma{}_{\nu\mu}=\frac{1}{2}g^{\sigma\delta}\left(\partial_{\nu}g_{\delta\mu}+\partial_\mu g_{\nu\delta}-\partial_\delta g_{\nu\mu}\right).
	\end{equation}
One particular convenient feature in three dimensions is that one can (Hodge) dualize the spin connection in such a way that it has the same index structure as the vielbein. In terms of Lorentz indices this can be achieved by using the 3d Levi-Civita symbol in order to obtain
	\begin{equation}\label{Intro:DualizedSpinConn}
		\omega^a=\frac{1}{2}\epsilon^{abc}\omega_{bc}\quad\Leftrightarrow\quad\omega_{ab}=-\epsilon_{abc}\omega^c,
	\end{equation}
where $\epsilon^{012}=1$. It is exactly this dualization of the spin connection which makes it possible to combine the vielbein and the spin connection into a single gauge field as we will review later in Section~\ref{ch:ChernSimonsFormulationofGravity}.\\
Using the dualized spin connection one can write the associated curvature two-form $R^a$ as
	\begin{equation}
		R^a=\extd\omega^a+\frac{1}{2}\epsilon^a{}_{bc}\omega^b\wedge\omega^c,
	\end{equation}
and consequently the Einstein-Hilbert-Palatini action \eqref{eq:EHSecondOrderAction} in terms of these new (first order) variables as
	\begin{equation}\label{Intro:IEHP}
		I_{EHP}=\frac{1}{8\pi G}\int_{\mathcal{M}} \left[e_a\wedge R^a -\frac{\Lambda}{6}\epsilon_{abc}e^a\wedge e^b\wedge e^c\right].
	\end{equation}
The equations of motion of the second order action \eqref{eq:EHSecondOrderAction} which are obtained by varying the action with respect to the metric $g_{\mu\nu}$ are given by the Einstein equations \eqref{eq:EinsteinEquationsSecondOrder}. Since in the frame-like formalism one has two independent fields $e^a$ and $\omega^{a}$ one has to vary \eqref{Intro:IEHP} with respect to both of these fields and subsequently also obtains two equations which encode curvature and torsion respectively as
	\begin{subequations}\label{eq:VielbeinCurvatureAndTorsion}
	\begin{align}
		R^a=&\extd\omega^a+\frac{1}{2}\epsilon^a{}_{bc}\omega^b\wedge\omega^c=\frac{\Lambda}{2}\epsilon^a{}_{bc} e^b\wedge e^c,\\
		T^a=&\extd e^a+\epsilon^a{}_{bc}\omega^b\wedge e^c=0.
	\end{align}
	\end{subequations}
This basic knowledge of frame fields, spin connections and how to use those two fields to cast the second order Einstein-Hilbert action \eqref{eq:EHSecondOrderAction} into a first order form \eqref{Intro:IEHP} is already sufficient to be able to move on to the the next section, in which we will describe how to rewrite the Einstein-Hilbert-Palatini action \eqref{Intro:IEHP} as a Chern-Simons action.

\subsection{Gravity as a Chern-Simons Theory}\label{ch:ChernSimonsFormulationofGravity}

As described in the previous section, instead of using a second order formalism, where the fundamental field of the theory is the metric $g_{\mu\nu}$, it can for some purposes be more convenient to use a first order formalism where the fundamental fields of the theory are the vielbein $e$ and the spin connection $\omega$. In three dimensions one finds that the dreibein and dualized spin connection have the same index structure in their Lorentz indices. Thus, one can combine these two quantities into a single gauge field
	\begin{equation}\label{eq:RelationAVielbeinSpinConnection}
		\mathcal{A}\equiv e^aP_a+\omega^aJ_a,
	\end{equation}
where the generators $P_a$ and $J_a$ generate the following Lie algebra
	\begin{equation}\label{eq:AllThreeAlgebras}
		[P_a,P_b]=-\Lambda\epsilon_{abc}J^c,\quad[J_a,J_b]=\epsilon_{abc}J^c,\quad[J_a,P_b]=\epsilon_{abc}P^c.
	\end{equation}
	\begin{itemize}
		\item For $\Lambda>0$, i.e. de Sitter spacetimes this gauge algebra is $\mathfrak{so}(3,1)$.
		\item For $\Lambda=0$, i.e. flat spacetimes this gauge algebra is $\mathfrak{isl}(2,\mathbb{R})\sim\mathfrak{sl}(2,\mathbb{R})\oplus_s\mathbb{R}^3$.
		\item For $\Lambda<0$, i.e. Anti-de Sitter spacetimes this gauge algebra is\\
		${\mathfrak{so}(2,2)\sim\mathfrak{sl}(2,\mathbb{R})\oplus\mathfrak{sl}(2,\mathbb{R})}$.
	\end{itemize}

Witten showed in 1988 \cite{Witten:1988hc} that the Chern-Simons action \cite{Achucarro:1987vz}
	\begin{equation}\label{eq:ChernSimonsAction}
		S_{\textnormal{CS}}[\mathcal{A}]=\frac{k}{4\pi}\int_{\mathcal{M}}\left<\mathcal{A}\wedge\extd\mathcal{A}+\frac{2}{3}\mathcal{A}\wedge\mathcal{A}\wedge\mathcal{A}\right>,
	\end{equation}
defined on a three-dimensional manifold $\mathcal{M}=\Sigma\times\mathbb{R}$, where $\Sigma$ is a two-dimensional manifold  with the invariant bilinear form
	\begin{equation}\label{eq:Isl2RBilinFormJanP}
		\left<J_aP_b\right>=\eta_{ab},\quad\left<J_aJ_b\right>=\left<P_aP_b\right>=0,
	\end{equation}
is indeed equivalent (up to boundary terms) to the Einstein-Hilbert-Palatini action in the first order formalism for positive, negative and zero cosmological constant \eqref{Intro:IEHP}, provided one identifies the Chern-Simons level $k$ with Newton's constant $G$ in three dimensions as
	\begin{equation}
		k=\frac{1}{4G}.
	\end{equation}

\paragraph{Anti-de Sitter Spacetimes:}

 One particular convenient feature of spacetimes with negative cosmological constant $\Lambda\equiv-\frac{1}{\ell^2}<0$ where $\ell$ is called the AdS radius, is that in a Chern-Simons formulation the underlying gauge symmetry $\mathfrak{so}(2,2)$ is a direct sum of two copies of $\mathfrak{sl}(2,\mathbb{R})$. This split can be made explicit by introducing the generators
	\begin{equation}
		J^\pm_a=\frac{1}{2}\left(J_a\pm\ell P_a\right).
	\end{equation}
These new generators satisfy
	\begin{equation}
		\left[J^+_a,J^-_b\right]=0,\qquad\left[J^\pm_a,J^\pm_b\right]=\epsilon_{abc}J^{c\pm}.
	\end{equation}
One can explicitly realize this split via
	\begin{equation}
		J^+_a=
			\begin{pmatrix}
				T^a & 0 \\
				0 & 0
			\end{pmatrix},\quad
		J^-_a=
			\begin{pmatrix}
				0 & 0 \\
				0 & \bar{T}^a
			\end{pmatrix},
	\end{equation}
where both $T^a$ and $\bar{T}^a$ satisfy an $\mathfrak{sl}(2,\mathbb{R})$ algebra. From \eqref{eq:Isl2RBilinFormJanP} one can immediately see that 
	\begin{equation}\label{eq:TTbarInvBilform}
		\left<T_a,T_b\right>=\frac{\ell}{2}\eta_{ab},\quad\left<\bar{T}_a,\bar{T}_b\right>=-\frac{\ell}{2}\eta_{ab}.
	\end{equation}
The gauge field $\mathcal{A}$ can now be written as
	\begin{equation}\label{eq:AdS3GaugeFieldSplit}
		\mathcal{A}=
			\begin{pmatrix}
				\left(\omega^a+\frac{1}{\ell}e^a\right)T_a & 0 \\
				0 & \left(\omega^a-\frac{1}{\ell}e^a\right)\bar{T}_a
			\end{pmatrix}\equiv
			\begin{pmatrix}
				A^aT_a & 0 \\
				0 & \bar{A}^a\bar{T}_a
			\end{pmatrix}.
	\end{equation}
Thus, after implementing this explicit split of $\mathfrak{so}(2,2)$ into a direct sum of two copies of $\mathfrak{sl}(2,\mathbb{R})$, the Chern-Simons action \eqref{eq:ChernSimonsAction} also splits into two contributions
	\begin{equation}
		S_{\textnormal{EH}}^{\textnormal{AdS}}[A,\bar{A}]=S_{\textnormal{CS}}[A]+S_{\textnormal{CS}}[\bar{A}],
	\end{equation}
where the invariant bilinear forms appearing in the Chern-Simons action are given by \eqref{eq:TTbarInvBilform}. Since both $T^a$ and $\bar{T}^a$ satisfy an $\mathfrak{sl}(2,\mathbb{R})$ algebra it is usually practical to not distinguish between the two generators, i.e. setting $T^a=\bar{T}^a$. This in turn also means that the invariant bilinear form in both sectors will be the same. From \eqref{eq:TTbarInvBilform}, however, we know that the invariant bilinear form in both sectors should have opposite sign. This is not a real problem since this relative minus sign can be easily introduced by hand by not taking the sum, but rather the difference of the two Chern-Simons actions
	\begin{equation}\label{eq:ChernSimonsActionAAbar}
		S_{\textnormal{EH}}^{\textnormal{AdS}}=S_{\textnormal{CS}}[A]-S_{\textnormal{CS}}[\bar{A}].
	\end{equation}
As the factor of $\ell$ in \eqref{eq:TTbarInvBilform} only yields an overall factor of $\ell$ to the action \eqref{eq:ChernSimonsActionAAbar} one can also absorb this factor simply in the Chern-Simons level as
	\begin{equation}\label{eq:ChernSimonsLevelNewtonConstant}
		k=\frac{\ell}{4G}.
	\end{equation}
This form of the Chern-Simons connection \eqref{eq:ChernSimonsActionAAbar} is usually the one discussed in the literature on AdS holography in $2+1$ dimensions. The big advantage of this split into an unbarred and a barred part in the case of AdS holography is that usually one only has to explicitly calculate things for one of the two sectors, as the other sector works in complete analogy, up to possible overall minus signs.\\
Up to this point we have only presented the basics of the Chern-Simons formulation of gravity in $2+1$ dimensions but did not go into detail as to \emph{why} exactly this formulation is so convenient and powerful for the purpose of studying the holographic principle. Thus, we will spend the remainder of this part of the lecture notes explaining the benefits of using the Chern-Simons formulation.\\
Maybe the biggest advantage of this formalism using Chern-Simons gauge fields is that this allows one to use all the techniques and machinery which is familiar from ordinary gauge theories. One can for example use finite gauge transformations of the form
	\begin{equation}\label{eq:FiniteCSGaugeTransformations}
		\mathcal{A}\rightarrow g^{-1}\left(\tilde{\mathcal{A}}+\extd\right)g,
	\end{equation}
where $g$ is some element of the group $G$ which is generated by some Lie algebra $\mathfrak{g}$ and $\mathcal{A}\in\mathfrak{g}$ to bring the gauge field $\mathcal{A}$ into a form which is convenient for the given task at hand. One can use for example a special gauge which is very convenient in the asymptotic analysis of AdS and non-AdS spacetimes whereas another gauge will be more convenient when making the transition from AdS to flat space. Since the gauge transformations \eqref{eq:FiniteCSGaugeTransformations} are finite in contrast to infinitesimal gauge transformations generated by a gauge parameter $\xi$ as
	\begin{equation}\label{eq:InfinitesimalCSGaugeTransformations}
		\delta_\xi\mathcal{A}=\extd\xi+[\mathcal{A},\xi],
	\end{equation}
one has to be careful which finite gauge transformations actually leave the Chern-Simons action \eqref{eq:ChernSimonsAction} invariant. In general a finite gauge transformation \eqref{eq:FiniteCSGaugeTransformations} changes the Chern-Simons action \eqref{eq:ChernSimonsAction} as $S_{\textnormal{CS}}[\mathcal{A}]\rightarrow S_{\textnormal{CS}}[\tilde{\mathcal{A}}]+\delta S_{\textnormal{CS}}[\tilde{\mathcal{A}}]$ with \cite{Blagojevic:2002aa}
	\begin{equation}
		\delta S_{\textnormal{CS}}[\tilde{\mathcal{A}}]=-\frac{k}{12\pi}\int_\mathcal{M}\left<g^{-1}\extd g\wedge g^{-1}\extd g\wedge g^{-1}\extd g\right>-\frac{k}{4\pi}\int_{\partial\mathcal{M}}\left<\extd gg^{-1}\wedge\tilde{\mathcal{A}}\right>.
	\end{equation}
This term vanishes for infinitesimal gauge transformations \eqref{eq:InfinitesimalCSGaugeTransformations} with gauge parameters $\xi\in\mathfrak{g}$ which are continuously connected to the identity $g\sim\unity+\xi$ and for finite gauge transformations which approach $g\rightarrow\unity$ sufficiently fast when approaching the boundary, but not for general finite gauge transformations\footnote{We will make this statement a bit more precise in Section~\ref{sec:3DCSEXample}.}. This means that there are finite gauge transformations of the form \eqref{eq:FiniteCSGaugeTransformations} which can change the state of the system and thus map between physically distinct setups.\\
Now considering the variation of \eqref{eq:ChernSimonsAction} with respect to the gauge field $\mathcal{A}$ one obtains the equations of motion of the Chern-Simons action \eqref{eq:ChernSimonsAction} as
	\begin{equation}
		F=\extd\mathcal{A}+\mathcal{A}\wedge\mathcal{A}=0,
	\end{equation}
which means that on-shell the Chern-Simons connection has to be locally flat. Remembering that the connection $\mathcal{A}$ can also be expressed in terms of a vielbein and spin connection as in \eqref{eq:RelationAVielbeinSpinConnection}, then requiring a flat connection $\mathcal{A}$ is equivalent to the equations \eqref{eq:VielbeinCurvatureAndTorsion}, which encode curvature and torsion. This is another check that the Chern-Simons action indeed correctly describes gravity in $2+1$ dimensions.\\
To require that the connection is locally flat also means that $\mathcal{A}=0$ is always a (trivial) solution of the equations of motion. Keeping in mind that finite gauge transformations in general can change the physical state, this in turn also means that for some holographic applications it can be beneficial to first start with the trivial configuration $\mathcal{A}=0$ and then use a finite gauge transformation \eqref{eq:FiniteCSGaugeTransformations} in order to obtain the desired result of a non-trivial configuration.\\
At this point we will also briefly elaborate on an important point of three-dimensional gravity, namely how diffeomorphisms appear in this gauge theoretic formulation. First consider the infinitesimal gauge transformation \eqref{eq:InfinitesimalCSGaugeTransformations} but now with a special gauge parameter of the form $\xi=\zeta^\nu \mathcal{A}_\nu$. After using the Leibniz rule one obtains
	\begin{equation}\label{eq:ChernSimonsFormulationDiffeosgaugeTrafos1}
		\delta_{(\zeta^\nu \mathcal{A}_\nu)}\mathcal{A}_\mu=\partial_\mu\zeta^\nu \mathcal{A}_\nu+\zeta^\nu\partial_\mu\mathcal{A}_\nu+\zeta^\nu[\mathcal{A}_\mu,\mathcal{A}_\nu].
	\end{equation}
Now adding $\zeta^\nu\left(\partial_\nu\mathcal{A}_\mu-\partial_\nu\mathcal{A}_\mu\right)$ to the right hand side of this equation does not really change anything. However, it allows one to rewrite \eqref{eq:ChernSimonsFormulationDiffeosgaugeTrafos1} in a more suggestive form as
	\begin{equation}
		\delta_{(\zeta^\nu \mathcal{A}_\nu)}\mathcal{A}_\mu=\mathcal{L}_{\zeta}\mathcal{A}_\mu+\zeta^\nu F_{\mu\nu},
	\end{equation}
where $\mathsterling_{\zeta}\mathcal{A}_\mu$ is the Lie derivative of the gauge field $\mathcal{A}_\mu$ given by
	\begin{equation} 
		\mathsterling_{\zeta}\mathcal{A}_\mu=\zeta^\nu\partial_\nu\mathcal{A}_\mu+\mathcal{A}_\nu\partial_\mu\zeta^\nu.
	\end{equation}
Thus, one can see that diffeomorphisms in three-dimensional gravity are on-shell (i.e. for $F=0$) equivalent to infinitesimal gauge transformations with gauge parameter $\xi=\zeta^\nu \mathcal{A}_\nu$. 

\subsection*{Non-AdS Spacetimes And Boundary Terms}

Since all of the interesting physics, aside global properties, in three-dimensional gravity are governed by degrees of freedom at the boundary it is of utmost importance to make sure that one can impose consistently fall off conditions of the gauge field\footnote{Or the metric in a second order formulation.} at the asymptotic boundary. Consistent in this context means that one still has a well defined variational principle after imposing said boundary conditions. This is crucial since a consistent variational principle is the core principle underlying the definition of equations of motion of a physical system described by some action. Thus the necessity of having such a well defined variational principle in turn also influences the possible set of boundary conditions that can be consistently imposed.\\
In order to see this let us take a closer look at the variation of the Chern-Simons action \eqref{eq:ChernSimonsAction}
	\begin{equation}
		\delta S_{\textrm{CS}}[\mathcal{A}]=\frac{k}{2\pi}\int_{\mathcal{M}}\left\langle\delta \mathcal{A}\wedge F\right\rangle+\frac{k}{4\pi}\int_{\partial\mathcal{M}}\left\langle\delta \mathcal{A}\wedge \mathcal{A}\right\rangle.
	\end{equation}
This expression only vanishes on-shell i.e. when $F=0$ if the second term on the right hand side vanishes as well. Assuming that the boundary $\partial\mathcal{M}$ is parametrized by a timelike coordinate $t$ and an angular coordinate $\varphi$ this amounts to
	\begin{equation}
		\frac{k}{4\pi}\int_{\partial\mathcal{M}}\left\langle\delta \mathcal{A}_t\mathcal{A}_\varphi-\delta\mathcal{A}_\varphi\mathcal{A}_t\right\rangle.
	\end{equation}
This term only vanishes if either $\mathcal{A}_\varphi$ or $\mathcal{A}_t$ are equal to zero everywhere. This is quite a stringent condition on possible boundary conditions and it would be nice to have a way of enlarging the possible set of consistent boundary conditions. This can be most easily done by simply adding a boundary term $B[\mathcal{A}]$ to the Chern-Simons action \eqref{eq:ChernSimonsAction}.\\
One could consider for example the following boundary term
	\begin{equation}
		B[\mathcal{A}]=\frac{k}{4\pi}\int_{\partial\mathcal{M}}\left\langle\mathcal{A}_\varphi \mathcal{A}_t\right\rangle.
	\end{equation}
Including this boundary term the total variation of the resulting action is on-shell
	\begin{equation}
		\delta S_{\textrm{CS}}[\mathcal{A}]^{\textrm{Tot}}=\frac{k}{2\pi}\int_{\partial\mathcal{M}}\left\langle\delta \mathcal{A}_t\mathcal{A}_\varphi\right\rangle.
	\end{equation}
Vanishing of the total variation then can be achieved either via
	\begin{equation}
		\mathcal{A}_\varphi\Bigr|_{\partial\mathcal{M}}=0\quad\textnormal{or}\quad\delta \mathcal{A}_t\Bigr|_{\partial\mathcal{M}}=0.
	\end{equation}
Choosing $\delta \mathcal{A}_t\Bigr|_{\partial\mathcal{M}}=0$, we are thus able to enlarge the possible set of boundary conditions by only having to making sure that the \emph{variation} of a part of the Chern-Simons connection has to vanish. One example where adding a boundary term\footnote{See \cite{Gary:2012ms} for more details on this.} is necessary is when one wants to describe spacetimes which fall into a class of so called non-AdS spacetimes. Such spacetimes are for example: null warped AdS, and their generalization Schr{\"o}dinger spacetimes \cite{Son:2008ye,Balasubramanian:2008dm,Adams:2008wt}, Lifshitz spacetimes, which are the gravity duals of Lifshitz-like fixed points \cite{Kachru:2008yh}, and the AdS/log CFT correspondence \cite{Grumiller:2008qz,Grumiller:2013at}.

\section{Canonical Analysis and Asymptotic Symmetries}

This section will be the most important one for what will follow as we will review the concept of quantizing gauge theories and give an explicit example on how this can be employed for gravity theories in three dimensions. Since there exist already excellent books such as e.g. \cite{Henneaux:1992,Blagojevic:2002aa} on how to quantize gauge systems we will keep the following introduction into the subject very short and focused on the aspects of the subject that are of most interest for the remainder of these lecture notes.

\subsection{General Ideas of Quantizing Gauge Theories}

In this section we will set the stage for the basic understanding of quantizing gauge theories that will be crucial for understanding the explicit example of 3D Chern-Simons theory that will be the follow up of this section.\\
Fundamental theories in physics tend to be gauge theories. That is, these theories contain physically redundant parameters in order to make the description of these systems more apparent. The trade off for such a description is that these additional parameters usually lead to a new symmetry that is called a gauge symmetry. These gauge transformations then transform two physically identically systems into each other. In addition these gauge symmetries are an important tool to extract the physical relevant information from the irrelevant ones as physical observables have to be invariant under gauge transformations. If a gauge transformation transforms two physically equivalent systems into each other one can also not expect that the equations of motion uniquely fix the time evolution of a gauge system because one can apply a gauge transformation at any given time. Hence it is a key property of gauge theories that general solutions of the equations of motion contain arbitrary functions of time.\\
As such the best way to treat gauge theories is via a Hamiltonian formulation. One of the main points of treating gauge theories as Hamiltonian systems is that the presence of arbitrary functions of time in general solutions of the equations of motion means that not all canonical variables are independent. As such a gauge system is always a constrained Hamiltonian system\footnote{It is important to note that the converse statement is not true.}.

\paragraph{The Lagrangian and Primary Constraints:}

Before we make the transition to a Hamiltonian description of gauge systems we want to start with an action principle in a Lagrangian formulation. Consider the action
    \begin{equation}
        S_{\mathcal{L}}=\int_{t_1}^{t_2}\mathcal{L}(q,\dot{q})\extd t,
    \end{equation}
where the Lagrangian $\mathcal{L}$ is a function of the coordinates $q^n$ and velocities $\dot{q}^n$ with $n=1,2,\ldots,N$. The equations of motion are determined by minimizing the functional for variations of the coordinates $\delta q^n$ that vanish at $t_1$ and $t_2$, i.e. $\delta q^n(t_1)=\delta q^n(t_2)=0$. This yields the well known Euler-Lagrange equations of motion
    \begin{equation}
       \frac{\extd}{\extd t}\left(\frac{\partial\mathcal{L}}{\partial\dot{q}^n}\right)-\frac{\partial\mathcal{L}}{\partial q^n}=0. 
    \end{equation}
More explicitly these equations can also be written as
    \begin{equation}
        \ddot{q}^m\frac{\partial^2\mathcal{L}}{\partial\dot{q}^m\partial\dot{q}^n}=\frac{\partial\mathcal{L}}{\partial q^n}-\dot{q}^m\frac{\partial^2\mathcal{L}}{\partial q^m\partial\dot{q}^n}.
    \end{equation}
This means that if the matrix
    \begin{equation}\label{eq:LagrangeJacobian}
        \frac{\partial^2\mathcal{L}}{\partial\dot{q}^m\partial\dot{q}^n},
    \end{equation}
is invertible i.e. the determinant does not vanish then the accelerations can be uniquely determined in terms of the positions and velocities. However, if the determinant of this matrix vanishes then that means that the accelerations are not uniquely determined in terms of positions and velocities and the equations of motion can contain arbitrary functions of time. This means that for gauge systems one is interested in the case where the determinant vanishes.\\
Departing from a Lagrangian formulation to a Hamiltonian one involves a Legendre transformation as well as the canonical momenta defined as
    \begin{equation}\label{eq:DefCanMomenta}
        p_n:=\frac{\partial\mathcal{L}}{\partial\dot{q}^n}.
    \end{equation}
Looking at \eqref{eq:LagrangeJacobian} as well as the definition of the canonical momenta one sees that the vanishing of the determinant of \eqref{eq:LagrangeJacobian} implies non-invertibility of the velocities as functions of positions and momenta. Thus not all momenta are independent from each other but rather satisfy some relations
    \begin{equation}
        \phi(p,q)_m=0,
    \end{equation}
that follow from the definition \eqref{eq:DefCanMomenta} of the canonical momenta. These relations are called \textbf{primary constraints} to highlight the fact that these relations do not involve the equations of motion and that they do not imply any restrictions on the positions $q$ and velocities $\dot{q}$.

\paragraph{The Canonical Hamiltonian:}

After having defined the canonical momenta we are ready to perform the Legendre transformation from the Lagrangian to the canonical Hamiltonian via
    \begin{equation}\label{eq:CanonicalHamiltonian}
        \mathcal{H}=\dot{q}^np_n-\mathcal{L}.
    \end{equation}
Looking at \eqref{eq:CanonicalHamiltonian} one might infer that $\mathcal{H}$ is a function of the momenta and velocities. However, looking at arbitrary variations of $\mathcal{H}$ 
    \begin{equation}
        \delta \mathcal{H}=\dot{q}^n\delta p_n+\delta\dot{q}^np_n-\delta\dot{q}^n\frac{\delta\mathcal{L}}{\delta\dot{q}^n}-\delta q^n\frac{\delta\mathcal{L}}{\delta q^n}=\dot{q}^n\delta p_n-\delta q^n\frac{\delta\mathcal{L}}{\delta q^n},
    \end{equation}
we see that the velocities $\dot{q}^n$ only enter via a very specific combination, that is the combination that gives the canonical momenta $p_n$. As such, the Hamiltonian is purely a function of the canonical momenta and positions.\\
However, it is important to note that the Hamiltonian \eqref{eq:CanonicalHamiltonian} is not uniquely determined in terms of the positions and canonical momenta as the canonical momenta have to satisfy the primary conditions $\phi_m=0$. This in turn means that the Hamiltonian is initially only properly defined on the constraint surface $\phi_m=0$ and has to be extended from that surface. From this it also follows that the formalism should be invariant under the change
    \begin{equation}
        \mathcal{H}\rightarrow \mathcal{H}+u^m\phi_m.
    \end{equation}
Using this Hamiltonian we can now also determine the time evolution of arbitrary functions of the canonical variables $F(p,q)$ via
    \begin{equation}\label{eq:TimeEvolution}
        \dot{F}=\{F,\mathcal{H}\}+u^m\{F,\phi_m\},
    \end{equation}
where the Poisson bracket $\{\cdot,\cdot\}$ is defined as
    \begin{equation}
        \{F,G\}=\frac{\partial F}{\partial q^n}\frac{\partial G}{\partial p_n}-\frac{\partial F}{\partial p_n}\frac{\partial G}{\partial q^n}.
    \end{equation}

\paragraph{Secondary Constraints and the Total Hamiltonian:}

The concept of secondary constraints is a direct consequence of the time evolution \eqref{eq:TimeEvolution}. That is that the primary constraints $\phi_m$ should be conserved in time i.e. $\dot{\phi}_m=0$. This can either lead to new relations among the canonical variables $p$ and $q$ or give restrictions on the parameters $u^m$. If the first case leads to restrictions that are independent of the primary constraints then these new restrictions are called \emph{secondary constraints}. Of course the secondary constraints should also preserved in time and this typically leads to a series of additional constraints which are usually all called secondary constraints to highlight the fact that in contrast to primary constraints the equations of motion were used to obtain these constraints.\\
After having determined all the secondary constraints the time evolution of all these constraints then typically leads to restrictions on the Lagrange multipliers $u^m$ which are given by\footnote{At this point it makes sense to introduce the weak equality symbol "$\approx$" to emphasize that a given quantity is numerically restricted to be zero but does not identically vanish on the whole phase space. Thus constraints are usually written as $\phi_n\equiv0$.}
    \begin{equation}\label{eq:LagrangeMultipliersEq}
        \{\phi_n,\mathcal{H}\}+u^m\{\phi_n,\phi_m\}\approx0,
    \end{equation}
where $\phi_n$ now is the collection of all primary and secondary constraints. Solving these equations one finds that the Lagrange multipliers $u^m$ can be written as $u^m=U^m+v^aV^m_a$, where $U^m$ is a particular solution of the inhomogeneous equation \eqref{eq:LagrangeMultipliersEq}, $V^m_a$ are linearly independent solutions of the most general solution of the homogeneous part of \eqref{eq:LagrangeMultipliersEq} and $\nu^a$ are completely arbitrary parameters.\\
This allows one to write down the \emph{total} Hamiltonian $\mathcal{H}_{\textrm T}$ as
    \begin{equation}
        \mathcal{H}_{\textrm T}=\mathcal{H}+U^m\phi_m+\nu^aV^m_a\phi_m.
    \end{equation}
This is the full Hamiltonian taking into account all constraints as well as their time evolution. Using this Hamiltonian the equations of motion are then simply given by
    \begin{equation}
        \dot{F}=\{F,\mathcal{H}_{\textrm T}\}.
    \end{equation}
    
\paragraph{First Class and Second Class Constraints:}

The distinction between primary and secondary constraints does not really have any influence on the description of gauge theories as constrained Hamiltonian systems. There is, however, a second more important distinction of constraints and that is whether or not constraints are \emph{first class} or \emph{second class}.\\
More generally a phase space function $F(p,q)$ is called first class if its Poisson bracket with every other constraint vanishes weakly i.e.
    \begin{equation}
        \{F,\phi_n\}\approx0.
    \end{equation}
A function that is not first class is called second class.\\
The distinction between first class and second class constraints is important since both have very different roles when it comes to quantizing a constrained Hamiltonian system. First class constraints are generators of gauge transformations and as such one has to very careful when quantizing a system\footnote{For a proof of that statement we refer the interested reader to \cite{Henneaux:1992}.}. Second class constraints on the other hand do not generate gauge transformations and can usually be strongly set to zero after introducing a modified Poisson bracket that is called a Dirac bracket. The logic here is the following: First, one uses the Poisson bracket to determine which constraints are first class and which ones are second class. Then after setting the second class constraints strongly equal to zero one discards the Poisson bracket in favor of the Dirac bracket. This change from Poisson to Dirac bracket is an essential step in quantizing the canonical commutation relations of the canonical variables of a constrained Hamiltonian system. If one would just blindly quantize the Poisson brackets of the canonical variables in the presence of second class constraints one would immediately encounter inconsistencies. However, the Dirac bracket correctly takes into account the second class constraints and thus allows for a consistent quantization.

\subsection{3D Chern-Simons Theory as an Explicit Example}\label{sec:3DCSEXample}

In this section we will use 3D Chern-Simons theory as an example to illustrate the principles explained in the previous section. Since there already exist excellent books explaining the basics of constrained Hamiltonian systems and canonical analysis, we also want to refer the interested reader to \cite{Henneaux:1992,Blagojevic:2002aa} for example.\\
The Chern-Simons gauge field $\mathcal{A}$ is a Lie algebra valued 1-form that can be written as
	\begin{equation}
		\mathcal{A}=\mathcal{A}^a{}_\mu\extd x^\mu T_a,
	\end{equation}
with $T_a$ being a basis of the Lie algebra $\mathfrak{g}$ one is considering. If one chooses such a basis then $\kappa_{ab}=\left\langle T_aT_b\right\rangle$ is a non-degenerate bilinear form on the Lie algebra. In components one can write \eqref{eq:ChernSimonsAction} as
	\begin{equation}\label{Intro:CoordAction}
		S_{\textrm{CS}}[\mathcal{A}]=\frac{k}{4\pi}\int_{\mathcal{M}}\extd^3x\epsilon^{\mu\nu\lambda}\kappa_{ab}\left(
		\mathcal{A}^a{}_\mu\partial_\nu \mathcal{A}^b{}_\lambda+\frac{1}{3}f^a{}_{cd}\mathcal{A}^c {}_\mu \mathcal{A}^d{}_\nu \mathcal{A}^b{}_\lambda\right),
	\end{equation}
where $\epsilon^{t\rho\varphi}=1$ and $f^a{}_{bc}$ are the structure constants of the Lie algebra given by
	\begin{equation}
		\left[T_a,T_b\right]=f^c{}_{ab}T_c.
	\end{equation}
Lie algebra indices $(a,b,\ldots)$ are raised and lowered with $\kappa_{ab}$ and spacetime indices $(\mu,\nu,\ldots)$ with the background metric $g_{\mu\nu}$ of the spacetime considered.\\
Proceeding with the canonical analysis it is convenient to use a $2+1$ decomposition of the action \eqref{eq:ChernSimonsAction} \cite{Banados:1994tn} that is given by
	\begin{equation}\label{Intro:2+1}
		S_{\textrm{CS}}[\mathcal{A}]=\frac{k}{4\pi}\int_{\mathbb{R}}\extd t\int_{\Sigma}\extd^2x\epsilon^{ij}\kappa_{ab}\left(\dot{\mathcal{A}}^a{}_i\mathcal{A}^b{}_j+\mathcal{A}^a{}_0F^b{}_{ij}+\partial_j
			\left(\mathcal{A}^a{}_i\mathcal{A}^b{}_0\right)\right),
	\end{equation}
with $F^a{}_{ij}=\partial_i\mathcal{A}^a{}_j-\partial_j\mathcal{A}^a{}_i+f^a{}_{bc}\mathcal{A}^b{}_i\mathcal{A}^c{}_j$ and $\epsilon^{ij}=\epsilon^{tij}$. Since the equations of motion require $F^a {}_{ij}=0$, the form of \eqref{Intro:2+1} already specifies $\mathcal{A}^a_0$ as a Lagrange multiplier and $\mathcal{A}^a_i$ as the dynamical fields. The Lagrangian density $\mathcal{L}$ is then given by
	\begin{equation}
		\mathcal{L}=\frac{k}{4\pi}\epsilon^{ij}\kappa_{ab}\left(\dot{\mathcal{A}}^a{}_i\mathcal{A}^b{}_j+\mathcal{A}^a{}_0F^b{}_{ij}+\partial_j\left(\mathcal{A}^a{}_i\mathcal{A}^b{}_0\right)\right).
	\end{equation}
Calculating the canonical momenta $\pi_a{}^\mu\equiv\frac{\partial\mathcal{L}}{\partial\dot{\mathcal{A}}^a_\mu}$ corresponding to the canonical variables $\mathcal{A}^a_\mu$ one finds the following primary constraints
	\begin{equation}
		\phi_a{}^0:=\pi_a{}^0\approx0\quad\phi_a{}^i:=\pi_a{}^i-\frac{k}{4\pi}\epsilon^{ij}\kappa_{ab}\mathcal{A}^b{}_j\approx0.
	\end{equation}
The Poisson brackets of the canonical variables are given by
	\begin{equation}\label{Intro:CanComm}
		\{\mathcal{A}^a{}_\mu(\textbf{x}),\pi_b{}^\nu(\textbf{y})\}=\delta^a{}_b\delta_\mu{}^\nu\delta^2(\textbf{x}-\textbf{y}).
	\end{equation}
The next step is to calculate the canonical Hamiltonian density via the following Legendre transformation 
	\begin{equation}
		\mathcal{H}=\pi_a{}^\mu\dot{\mathcal{A}}^a{}_\mu-\mathcal{L}=-\frac{k}{4\pi}\epsilon^{ij}\kappa_{ab}\left(\mathcal{A}^a{}_0F^b{}_{ij}+\partial_j\left(\mathcal{A}^a{}_i\mathcal{A}^b{}_0\right)\right).
	\end{equation}
Since we are dealing with a constrained Hamiltonian system, we have to work with the total Hamiltonian given by
	\begin{equation}
		\mathcal{H}_T=\mathcal{H}+u^a{}_\mu\phi_a{}^\mu,
	\end{equation} 
where $u^a{}_\mu$ are some arbitrary multipliers. Since the primary constraints should be conserved after a time evolution, we require
	\begin{equation}
		\dot{\phi}_a{}^\mu=\{\phi_a{}^\mu,\mathcal{H}_T\}\approx0,
	\end{equation}
which leads to the following secondary constraints
	\begin{align}
		\mathcal{K}_a\equiv-\frac{k}{4\pi}\epsilon^{ij}\kappa_{ab}F^b{}_{ij}&\approx0\\
		D_i\mathcal{A}^a{}_0-u^a{}_i&\approx0,\label{Intro:Multiplier}
	\end{align}
where $D_iX^a=\partial_iX^a+f^a{}_{bc}\mathcal{A}^b{}_iX^c$ is the gauge covariant derivative. One can now use the Hamilton equations of motion, which are given by
	\begin{equation}
		\dot{\mathcal{A}}^a{}_i=\frac{\partial\mathcal{H}_T}{\partial\pi_a{}^i}=u^a{}_i
	\end{equation}
to determine the Lagrange multipliers $u^a{}_i$ and rewrite \eqref{Intro:Multiplier}. This yields the following weak equality
	\begin{equation}
		D_i\mathcal{A}^a{}_0-u^a{}_i = D_i\mathcal{A}^a{}_0-\partial_0\mathcal{A}^a{}_i = F^a{}_{i0}\approx0.
	\end{equation}
The total Hamiltonian can now be written in the following form
	\begin{equation}
		\mathcal{H}_T=\mathcal{A}^a{}_0\bar{\mathcal{K}}_a+u^a{}_0\phi_a{}^0+\partial_i(\mathcal{A}^a{}_0\pi_a{}^i),
	\end{equation}
with
	\begin{equation}
		\bar{\mathcal{K}}_a=\mathcal{K}_a-D_i\phi_a{}^i.
	\end{equation}
One can use the canonical commutation relations \eqref{Intro:CanComm} to determine the following Poisson brackets which will be necessary to determine the Poisson algebra of the constraints
	\begin{subequations}\label{Intro:ConstraintBrackets}
		\begin{align}
			\{\phi_a{}^0(\textbf{x}),\mathcal{A}^b{}_0(\textbf{y})\}&=-\delta_a{}^b\delta^2(\textbf{x}-\textbf{y}),\\
			\{\phi_a{}^i(\textbf{x}),\mathcal{A}^b{}_j(\textbf{y})\}&=-\delta_a{}^b\delta^i{}_j\delta^2(\textbf{x}-\textbf{y}),\\
			\{\phi_a{}^i(\textbf{x}),\pi_b{}^j(\textbf{y})\}&=-\frac{k}{4\pi}\epsilon^{ij}\kappa_{ab}\delta^2(\textbf{x}-\textbf{y}),\\
			\{\phi_a{}^i(\textbf{x}),\pi_b{}^j(\textbf{y})\}&=-\frac{k}{2\pi}\epsilon^{ij}\kappa_{ab}\delta^2(\textbf{x}-\textbf{y}),\\
			\{\mathcal{A}^a{}_i(\textbf{x}),D_j\phi_b{}^j(\textbf{y})\}&=[\delta^a{}_b\partial_i+f^a{}_{bc}\mathcal{A}^c{}_i(\textbf{y})]\delta^2(\textbf{x}-\textbf{y}),\\
			\{\pi_a{}^i(\textbf{x}),D_j\phi_b{}^j(\textbf{y})\}&=-\frac{k}{4\pi}\epsilon^{ij}[\kappa_{ab}\partial_j+f_{abc}\mathcal{A}^c{}_j(\textbf{y})]\delta^2(\textbf{x}-\textbf{y})
										+f_{ab}{}^c\phi_c{}^i(\textbf{y})\delta^2(\textbf{x}-\textbf{y}),\\
			\{\phi_a{}^i(\textbf{x}),D_j\phi_b{}^j(\textbf{y})\}&=-\frac{k}{2\pi}\epsilon^{ij}[\kappa_{ab}\partial_j+f_{abc}\mathcal{A}^c{}_j(\textbf{y})]\delta^2(\textbf{x}-\textbf{y})
										+f_{ab}{}^c\phi_c{}^i(\textbf{y})\delta^2(\textbf{x}-\textbf{y}),\\
			\{\pi_a{}^i(\textbf{x}),\mathcal{K}_b(\textbf{y})\}&=-\frac{k}{2\pi}\epsilon^{ij}[\kappa_{ab}\partial_j+f_{abc}\mathcal{A}^c{}_j(\textbf{y})]\delta^2(\textbf{x}-\textbf{y}),\\
			\{\phi_a{}^i(\textbf{x}),\mathcal{K}_b(\textbf{y})\}&=-\frac{k}{2\pi}\epsilon^{ij}[\kappa_{ab}\partial_j+f_{abc}\mathcal{A}^c{}_j(\textbf{y})]\delta^2(\textbf{x}-\textbf{y}),\\
			\{D_i\phi_a{}^i(\textbf{x}),\mathcal{K}_b(\textbf{y})\}&=-\frac{k}{2\pi}\epsilon^{ij}f_{abc}D_i\mathcal{A}^c{}_j\delta^2(\textbf{x}-\textbf{y}),\\
			\{\phi_a{}^i(\textbf{x}),\bar{\mathcal{K}}_b(\textbf{y})\}&=-f_{ab}{}^c\phi_c{}^i\delta^2(\textbf{x}-\textbf{y}),\\
			\{D_i\phi_a{}^i(\textbf{x}),D_j\phi_b{}^j(\textbf{y})\}&=-\frac{k}{2\pi}\epsilon^{ij}f_{abc}D_i\mathcal{A}^c{}_j\delta^2(\textbf{x}-\textbf{y})-
											f_{ab}{}^cD_i\phi_c{}^i\delta^2(\textbf{x}-\textbf{y}),
		\end{align}
	\end{subequations}
where $\partial_i$ denotes $\frac{\partial}{\partial y^i}$. Using these relations one finds the following algebra of constraints
	\begin{subequations}
		\begin{align}
			\{\phi_a{}^i(\textbf{x}),\phi_b{}^j(\textbf{y})\}&=-\frac{k}{2\pi}\epsilon^{ij}\kappa_{ab}\delta^2(\textbf{x}-\textbf{y}),\\
			\{\phi_a{}^i(\textbf{x}),\bar{\mathcal{K}}_b(\textbf{y})\}&=-f_{ab}{}^c\phi_c{}^i\delta^2(\textbf{x}-\textbf{y}),\\
			\{\bar{\mathcal{K}}_a(\textbf{x}),\bar{\mathcal{K}}_b(\textbf{y})\}&=-f_{ab}{}^c\bar{\mathcal{K}}_c\delta^2(\textbf{x}-\textbf{y}),
		\end{align}
	\end{subequations}
which are the only non-vanishing Poisson brackets of the constraints $\phi_a{}^\mu$ and $\bar{\mathcal{K}}_a$. Hence $\phi_a{}^0$ and $\bar{\mathcal{K}}_a$ are first class constraints and $\phi_a{}^i$ are second class constraints. Thus we can use the second class constraints $\phi_a{}^i$ to restrict our phase space and define the corresponding Dirac bracket of the remaining canonical variables. In this case the only non-vanishing Dirac bracket of the dynamical fields is given by the following relation
	\begin{equation}
		\{\mathcal{A}^a{}_i(\textbf{x}),\mathcal{A}^b{}_j(\textbf{y})\}_{\textrm{D.B.}}=\frac{2\pi}{k}\kappa^{ab}\epsilon_{ij}\delta^2(\textbf{x}-\textbf{y}).
	\end{equation}
As a next step we are interested in the generators that correspond to the gauge transformations induced by the first class constraints $\phi_a{}^0$ and $\bar{\mathcal{K}}_a$. A useful way to construct the generators is given by Castellani's algorithm \cite{Castellani:1981us}. In the general case the gauge generator is given by
	\begin{equation}
		G=\lambda(t)G_0+\dot{\lambda}(t)G_1,
	\end{equation}
with $\dot{\lambda}(t)\equiv\frac{d\lambda(t)}{dt}$. The constraints $G_0$ and $G_1$ then have to fulfill the following relations
	\begin{subequations}
		\begin{align}
			G_1&=C_{PFC},\\
			G_0+\{G_1,\mathcal{H}_T\}&=C_{PFC},\\
			\{G_0,\mathcal{H}_T\}&=C_{PFC},
		\end{align}
	\end{subequations}
where $C_{PFC}$ denotes a primary first class constraint. These relations are fulfilled for $G_0=\bar{\mathcal{K}}_a$ and $G_1=\phi_a{}^0=\pi_a{}^0$. The smeared generator of gauge transformations has the following form
	\begin{equation}
		G[\lambda]=\int_\Sigma\extd^2x\left(D_0\lambda^a\pi_a{}^0+\lambda^a\bar{\mathcal{K}}_a\right).
	\end{equation}
Using \eqref{Intro:ConstraintBrackets} one can show by a straightforward calculation that this generator generates the following gauge transformations via $\delta_\lambda\bullet=\{\bullet,G[\lambda]\}$
	\begin{subequations}
		\begin{align}
			\delta_\lambda \mathcal{A}^a{}_0&=D_0\lambda^a,\\
			\delta_\lambda \mathcal{A}^a{}_i&=D_i\lambda^a,\\
			\delta_\lambda\pi_a{}^0&=-f_{ab}{}^c\lambda^b\pi_c{}^0,\\
			\delta_\lambda\pi_a{}^i&=\frac{k}{4\pi}\epsilon^{ij}\kappa_{ab}\partial_j\lambda^b-f_{ab}{}^c\lambda^b\pi_c{}^i,\\
			\delta_\lambda\phi_a{}^i&=-f_{ab}{}^c\lambda^b\phi_c{}^i.
		\end{align}
	\end{subequations}
The generator $G$ that we have constructed via this method is only a preliminary result, since the presence of a boundary in our theory prevents  that the generator $G$ is properly functionally differentiable. We will fix this by first computing the full variation of the generator for a field independent gauge parameter $\lambda^a$
	\begin{align}\label{Intro:DeltaG}
		\delta G[\lambda]=&\int_\Sigma\extd^2x(\delta(D_0\lambda^a\pi_a{}^0)+\lambda^a\delta\bar{K}_a)=\nonumber\\
				      &\int_\Sigma\extd^2x\left(\dot{\lambda}^a\delta\pi_a{}^0-\lambda^af_{ab}{}^c(\delta \mathcal{A}^b{}_0\pi_c{}^0+\mathcal{A}^b{}_0\delta\pi_c{}^0)-\frac{k}{4\pi}
					\epsilon^{ij}\kappa_{ab}\partial_j\lambda^a\delta \mathcal{A}^b{}_i+\right.\nonumber\\
				      &\left.\partial_i\lambda^a\delta\pi_a{}^i-\lambda^af_{ab}{}^c(\delta \mathcal{A}^b{}_i\pi_c{}^i+\mathcal{A}^b{}_i\delta\pi_c{}^i)-
					\partial_i\left(\frac{k}{4\pi}\epsilon^{ij}\kappa_{ab}\lambda^a\delta \mathcal{A}^b{}_j+\lambda^a\delta\pi_a{}^i\right)\right)\nonumber\\
				   =&\int_\Sigma\extd^2x\left(f^a{}_{bc}\lambda^c\pi_a{}^\mu\delta \mathcal{A}^b{}_\mu+D_\mu\lambda^a\delta\pi_a{}^\mu+
					\frac{k}{4\pi}\epsilon^{ij}\kappa_{ab}\partial_i\lambda^a\delta \mathcal{A}^b{}_j-\right.\nonumber\\
				      &\left.\partial_i\left(\frac{k}{4\pi}\epsilon^{ij}\kappa_{ab}\lambda^a\delta \mathcal{A}^b{}_j+\lambda^a\delta\pi_a{}^i\right)\right).
	\end{align}
The first three terms are regular bulk terms and thus do not spoil functional differentiability. The last term on the other hand is a boundary term that spoils functional differentiability. Thus in order to fix this one has to add a suitable boundary term to the gauge generator such that the variation of this additional boundary term cancels exactly the boundary term in  \eqref{Intro:DeltaG} i.e.
	\begin{equation}
		\delta\bar{G}[\lambda]=\delta G[\lambda]+\delta Q[\lambda],
	\end{equation}
with
	\begin{equation}
		\delta Q[\lambda]=\int_\Sigma\extd^2x\,\partial_i\left(\frac{k}{4\pi}\epsilon^{ij}\kappa_{ab}\lambda^a\delta \mathcal{A}^b{}_j+\lambda^a\delta\pi_a{}^i\right).
	\end{equation}
Setting the second class constraints $\phi_a{}^i\approx0$ strongly equal to zero, thus going into the reduced phase space and using in addition Stoke's theorem, the variation of the boundary charge can be written as
	\begin{equation}
		\delta Q[\lambda]=\frac{k}{2\pi}\int\extd\varphi \kappa_{ab}\lambda^a\delta \mathcal{A}^b{}_\varphi.
	\end{equation}
If we assume that the gauge parameter is field independent, then the boundary charge $Q[\lambda]$ is trivially integrable. This yields the following canonical boundary charge
	\begin{equation}
		Q[\lambda]=\frac{k}{2\pi}\int\extd\varphi \kappa_{ab}\lambda^a\mathcal{A}^b{}_\varphi.
	\end{equation}
After performing the canonical analysis and having identified all the constraints we can turn our attention to an appropriate choice of gauge. Since we have found two first class constraints we are free to impose two sets of gauge conditions. One appropriate partial gauge fixing choice is given by \cite{Blagojevic:2002aa}
	\begin{subequations}\label{Intro:GaugeChoice}
		\begin{align}
			\mathcal{A}_\rho={}&b^{-1}(\rho)\partial_\rho b(\rho),\\
			\mathcal{A}_\varphi={}&b^{-1}(\rho)a_\varphi(\varphi,t)b(\rho),\\
			\mathcal{A}_t={}&b^{-1}(\rho)a_t(\varphi,t)b(\rho).
		\end{align}
	\end{subequations}
This choice of gauge automatically solves the flatness conditions $F_{t\rho}=0$ and $F_{\varphi\rho}=0$.\\
One possible choice of such a group element is given by	
	\begin{equation}\label{Intro:BChoice}
		b(\rho)=e ^{\rho L_0},
	\end{equation}
where $L_0\in\mathfrak{sl}(2,\mathbb{R})$ in a basis where
    \begin{equation}
        [L_n,L_m]=(n-m)L_{n+m},
    \end{equation}
for $n,m=\pm1,0$. This choice of $b(\rho)$ extensively used in the AdS$_3$ Chern-Simons gravity literature as it corresponds to a Fefferman-Graham expansion of the metric if translated properly to the second order formalism.

\subsection{Computing Charges in the Metric Formulation}

The Chern-Simons formalism is very powerful to compute easily the conserved charges\footnote{See e.g. \cite{Miskovic:2016mvs} where the Chern-Simons formalism is used to compute the vacuum energy of asymptotically flat spacetimes in three dimensions.}. But its field of applications is limited to three dimensions whereas the metric formulation can be used for an arbitrary number of dimensions. However, its downside is the heaviness of the computations.\\
Before briefly explaining a few famous methods used to compute conserved quantities in gravitation, we say a few words on why the computation of the charges is so peculiar for gauge symmetries. This introduction is based on \cite{Compere:2007az,Compere:2006my}.\\
In gravitation, there is no notion of a local energy-momentum tensor as the equivalence principle states that locally the effects of gravity can be always suppressed. Thus, we have to look for another type of quantity to define for example the energy of a spacetime.
It took quite a long time to realize this and the first answers to that problem were proposed in the early sixties by Arnowitt, Deser and Misner \cite{Arnowitt:1962hi}. Before looking at their method, we use a slightly more modern point of view to understand why the conserved quantities in gravity, and more generally in gauge theories, are not described in terms of volume charges but rather as surface charges.\\
The ordinary (or classical or first) Noether theorem is the following \cite{Barnich:1994db,Barnich:1995ap} (see the proof therein) 
\begin{theorem}Ordinary Noether theorem: Continuous global symmetries (defined up to gauge transformations) of a Lagrangian are in one-to-one correspondence with equivalence classes of conserved currents.\end{theorem} 
Two currents $J,J'$ are equivalent if they differ by a trivial current, the latter being constituted of a quantity vanishing on-shell and/or a gauge transformation, 
\begin{equation}
J^\mu= J'^\mu+ \partial_\nu k^{[\mu\nu]} + t^\mu(\textrm{EOM}), 
\end{equation}
with $t^\mu$ vanishing on-shell. 
The conserved current $J$ can be used to defined the conserved charge
\begin{equation}\label{truenoether}
Q=\int_\Sigma \extd^{n-1}x\, J^0,  
\end{equation}
where $\Sigma$ is a spacelike Cauchy surface\footnote{If $J$ is a conserved current, $\p_\mu J^\mu=0$, the charge Q is indeed conserved in time $\p_t Q=\int_\Sigma \extd^{n-1}x\,\p_t J^0=\int_{\Sigma}\extd^{n-1}x \,(-\p_iJ^i)=\int_{\p \Sigma} \extd^{n-2}x\,J^i=0$ if the fields decrease sufficiently fast at asymptotic infinity.}.\\
Before going further, we make a small remark on the notation of the so-called $(n-p)$-forms where $n$ is the dimension of the spacetime. The Hodge dual of a $p$-form $k$ is $\star k = k^{\nu_1...\nu_p}(\extd^{n-p}x)_{\nu_1...\nu_p} $ with 
\begin{equation}
(\extd^{n-p}x)_{\nu_1...\nu_p}=\frac1{(n-p)!p!}\epsilon_{\nu_1...\nu_p\nu_{p+1}...\nu_n}\extd x^{\nu_{p+1}}\wedge ... \wedge \extd x^{\nu_n}\,.
\end{equation}
With a slight abuse of notation, we call $k^{\nu_1...\nu_p}$ a $(n-p)$-form. These objects will be essential to describe conserved quantities for gauge symmetries.\\
Noether theorem implies that for gauge symmetries all the currents are trivial. Therefore, they cannot be used to define a non-trivial conserved charge.\\
Let us see this explicitly for the example of Maxwell theory. The action in this case is \begin{equation}
    S=-\frac14\int \extd^n x\, F^{\mu\nu}F_{\mu\nu}\,,
\end{equation}
whose equations of motions are $\p_\mu F^{\mu\nu}=0$ and which is invariant under the gauge symmetries
\begin{equation}\label{MaxGauge}
    \delta_s A_\mu= \p_\mu c(x) \,.
\end{equation}
The Noether current is defined as, for an action $S=\int \extd^n x \,L(\phi,\p_\mu \phi)$ and for $\delta_s \phi$ a symmetry of the theory, i.e $\delta_s L= \p_\mu K^\mu$,
\begin{equation}\label{Noethercurrentclass}
 J^\mu = - K^\mu +\frac{\p L}{\p\p_\mu\phi}\delta_s \phi \,. 
\end{equation}
For the symmetry \eqref{MaxGauge}, $K^\mu= 0$ and $\frac{\p L}{\p\p_\mu\phi}\delta_s \phi=-F^{\mu\nu}\p_\nu c(x) $. Thus the Noether current
\begin{equation}
    J^\nu= - F^{\mu\nu}\p_\mu c(x)=-\p_\mu( F^{\mu\nu} c(x)) + \p_\mu F^{\mu\nu} c(x),
\end{equation}
is trivial, being the sum of a divergence of an antisymmetric tensor and a term proportional to the equations of motion. Thus, it can not be used to define directly a charge. \\
However, gauge symmetries have conserved charges. They are the objects of the following theorem \cite{Barnich:1994db,Barnich:1995ap}:
\begin{theorem}Generalized Noether theorem: A reducibility parameter is in correspondence with a $(n-2)$-form conserved on-shell (up to trivial $(n-3)$-forms and up to the addition of the divergence of a $(n-3)$-form). \end{theorem}
A reducibility parameter is a parameter of a gauge transformation such that its associated gauge transformation vanishes on-shell but not the parameter itself. The conserved $(n-2)$-form is sometimes called superpotential in the literature. 
In the case of this theorem, the conserved charge is obtained by integrating the current over a surface at constant time and thus it is unsurprisingly called a surface charge. \\
As an example, let us consider the case of Maxwell theory whose gauge transformations are $\delta A_\mu=\p_\mu c(x)$. The reducibility parameter is such that its gauge transformation vanishes on-shell, $
\p_\mu c(x)=0$, and it is non trivial, i.e. $c(x)\neq 0$. It is trivially a non-zero constant. The theorem ensures the existence of a conserved $(n-2)$-form but it does not give a prescription to find it (see examples of it in section \ref{CovPhaseForm}). However, it can be easily guessed for Maxwell's theory as it has to depend on the reducibility parameter, be built from the matter content and be conserved. It is given by
\begin{equation}
k^{\mu\nu}=c\, F^{\mu\nu}\,.
\end{equation}
The electric charge is obtained by integrating this $(n-2)$-form over a spacelike $(n-2)$ surface $C$
\begin{equation}
    Q_{c=1}=\oint_{C}k\,.
\end{equation}
It is the same charge appearing in Gauss' law. \\
The case of gravitation does not enjoy the same simplicity. Indeed, it is not possible to solve the Killing equation $\nabla_\mu\xi_\nu+\nabla_\nu \xi_\mu=0$ for any spacetime. Thus, there is no general formula for a conserved $(n-2)$-form.\\
These considerations explain why a posteriori it took so much time to propose a expression of the energy of a spacetime. It is associated to the time translations but the latter is a diffeomorphism, i.e. gauge symmetry, for gravitational theories (as opposed to the case of pure Maxwell theory where this symmetry is global and where the first Noether theorem is used to define the energy).\\
Over the years, many ways were found to determine an expression for this conserved $(n-2)$-form in certain contexts. 
Not of all them are equivalent and are more or less easy to apply depending on the case of interest. 
In the next subsections, we present some of the most known techniques.

\subsubsection{Arnowitt Deser Misner (ADM) Formulation - Hamiltonian approach}

This first method was developed by Arnowitt, Deser and Misner in \cite{Arnowitt:1962hi} where they proposed a formula for the energy of asymptotically flat spacetimes in 3+1 dimensions. Their derivation is based on the Hamiltonian formalism of general relativity introduced previously in this notes, see also Appendix E of Wald's book \cite{Wald:106274}. 
The authors using this transcription to the Hamiltonian formalism were able to enlighten our comprehension of general relativity. Here, we focus very briefly on the computation of the energy. For more interesting features, see for example \cite{Arnowitt:1962hi,Wald:106274,Blau,Misner1973}. 
Recall that the Hamiltonian of general relativity is 
\begin{equation}
H=\int_{\Sigma} ( N\,\mathcal H + \mathcal N_i \mathcal H^i )+ \text{boundary terms} \,,
\end{equation}
where $N, \mathcal N_i$ are the lapse and shift functions and $i$ runs over spatial indices. Here they can be seen as Lagrange multipliers to ensure the so-called Hamiltonian and momentum constraints $\mathcal{H}=0$ and $\mathcal H^i=0$.\\
The energy of a physical state is given by the Hamiltonian. As the constraints are zero on a physical state, the energy is only given by boundary terms. 
We recover here the main idea that in gauge theory, there are no volume charges but rather surface charges. 
The precise expression for the boundary terms is rather complicated and can be found for example in \cite{Blau}.
These results were later extended by Regge and Teitelboim \cite{Regge:1974zd} to an expression for the momentum and angular momentum independently of the coordinates. 
Later, the Hamiltonian formalism was also successfully applied to anti-de Sitter spacetimes \cite{Henneaux:1985tv}. 

\subsubsection{Abbott Deser Tekin (ADT) Formalism}

Twenty years after ADM, Abbott and Deser \cite{Abbott:1981ff} derived an expression of the energy in general relativity for non-vanishing cosmological constant theories in four dimensions. Deser and Tekin extended it to any dimension, any theory with higher curvature terms and any constant curvature \cite{Deser:2002jk,Deser:2002rt}. Their method is based on the linearization of the theory around a constant curvature background. Let us present briefly the steps leading to the energy.\\
From Einstein's theory of gravity, the equations of motion \eqref{eq:EinsteinEquationsSecondOrder} can be derived and more compactly written as
\begin{equation}\label{ADTEOM}
\mathcal E_{\mu\nu}=\kappa \, \tau_{\mu\nu}\,,
\end{equation}
with $\kappa$ usually taken to be $8\pi G$ and $\tau_{\mu\nu}$ the energy momentum tensor of the matter fields. 
The metric is decomposed in the constant curvature background (solving $\mathcal E_{\mu\nu}(\bar g_{\mu\nu} )=0$) plus a perturbation (not necessarily small) vanishing sufficiently fast at infinity
\begin{equation}
g_{\mu\nu}=\bar g_{\mu\nu}+h_{\mu\nu}\,. 
\end{equation}
The bar notation always denotes a quantity written in terms of the background $\bar g_{\mu\nu}$. 
The equations of motion \eqref{ADTEOM} can be expanded in terms of the background and the perturbation. The contribution of 0$^{\textrm {th}}$ order in the perturbation vanishes by definition of the background metric. We write the rest as the linear part in the perturbation $\mathcal E^L_{\mu\nu}$ and an effective energy-momentum tensor $T_{\mu\nu}$, consisting of the matter contribution $\tau_{\mu\nu}$ and the non-linear terms in $h_{\mu\nu}$, 
\begin{equation}
\mathcal E^L_{\mu\nu}=\kappa \, T_{\mu\nu}\,. 
\end{equation}
Using the Bianchi identities of \eqref{ADTEOM}, 
$\nabla_\mu \mathcal E^{\mu\nu}=0$, it is straightforward to show that 
\begin{equation}
\bar{\nabla}_\mu \mathcal E^L_{\mu\nu}=0 \, , \text{ and therefore } \bar{\nabla}_\mu T^{\mu\nu}=0\,. 
\end{equation}
Be $\bar \xi$ a Killing vector of the background metric $\bar g$, 
$\bar \nabla_\mu\bar \xi_\nu+\bar \nabla_\nu\bar \xi_\mu=0\,,$
one can construct a conserved current $T^{\mu\nu}\bar\xi_\nu$ 
\begin{equation}
\bar \nabla_\mu(T^{\mu\nu}\bar\xi_\nu ) =\frac1{\sqrt{-\bar g}}\partial_\mu (\sqrt{-\bar g} T^{\mu\nu}\bar\xi_\nu ) =0 \,.
\end{equation}
Thus, it exists an antisymmetric tensor such that 
\begin{equation}
T^{\mu\nu}\bar\xi_\nu =\bar{\nabla}_\nu\mathcal F^{\mu\nu}\,. 
\end{equation}
Also the conserved charge associated to $\bar{\xi}$ is a surface charge
\begin{equation}
Q^\mu(\bar \xi)=\int_M \sqrt{-\bar g}  T^{\mu\nu}\bar\xi_\nu=\int_{\p M}\mathcal F^{\mu i }\extd S_i \,,
\end{equation}
where $M$ is a $(n-1)$-dimensional hypersurface and i ranges over $(1,...,n-2)$.\\
Now, the difficulty for each theory is to build this $\mathcal F$. For example, for general relativity the conserved charges are \cite{Deser:2002jk}
\begin{align} \nonumber
Q^\mu(\bar \xi)=\frac1{4 \Omega_{n-2}G_n}\int_{\partial M} \extd S_i\Big(&
\bar\xi_\nu\bar\nabla^\mu h^{i\nu}-\bar\xi_\nu\bar\nabla^i h^{\mu\nu} +\bar\xi_\mu\bar\nabla^i h -\bar\xi^i\bar\nabla^\mu h\\ 
&+h^{\mu\nu}\bar\nabla^i \bar\xi_\nu - h^{i\nu}\bar\nabla^\mu \bar\xi_\nu
+\bar\xi^i\bar\nabla_\nu h^{\mu\nu}-\bar\xi^\mu\bar\nabla_\nu h^{i\nu}
+h \bar\nabla^\mu \bar\xi^i 
\Big)\,, 
\end{align}
where $\Omega_{n-2}$ is the solid angle and $G_n$ the n-dimensional Newton constant.
	
\subsubsection{Covariant Phase Space Formalism}\label{CovPhaseForm}

We present briefly the covariant phase formalism for the case where only the metric is present and we do so for simplicity reasons. 
We first discuss the Iyer-Wald formalism \cite{Wald:1993nt,Iyer:1994ys}. After, we say a few words on the Brandt-Barnich-Comp{\`e}re method \cite{Barnich:2001jy, Barnich:2007bf}. See also for example the review in \cite{Detournay:2016gao}, section 3 of \cite{Compere:2009dp,Azeyanagi:2009wf}, App. A of \cite{Anninos:2011vd}.\\
From the $n$-form Lagrangian $L$, the equations of motion $E(g)=0$ are determined as
\begin{equation} \delta L (g) =E(g) \delta g+\extd\Theta(\delta g,g),\label{eq49}
\end{equation}
where $\Theta[\delta g,g]$ is the symplectic potential $(n-1)$-form. 
The gauge symmetries of the theory are transformations $\delta_\xi g = \mathsterling_\xi g$ under which the Lagrangian transforms as 
\begin{equation} \label{eq50}
  \delta_\xi L = \mathsterling_\xi L =( \extd \, i_\xi+i_\xi \,\extd\,)L=\extd \,(i_\xi L) 
  \,,
\end{equation}
where $i$ denotes the interior product. 
 On the other hand, writing (\ref{eq49}) for a gauge transformation ($\delta = \delta_\xi$) and using (\ref{eq50}), one gets 
\begin{equation} \label{eq52}
 E(g) \delta_\xi g= -\extd J_\xi(g),
\end{equation}
where the Noether current is defined as\footnote{It is the same definition as in \eqref{Noethercurrentclass} but in the dual picture. Indeed, the $\star K^\mu$ is given by $i_\xi L$. To link the $\Theta$ of Iyer-Wald with \eqref{Noethercurrentclass}, we take a Lagrangian $L(\phi,\p_\mu\phi)$. We have $\delta L= \frac{\p L}{\p \phi}\delta \phi +  \frac{\p L}{\p \p_\mu\phi}\delta \p_\mu\phi=  (\frac{\p L}{\p \phi}- \frac{\p L}{\p\p_\mu \phi})\delta \phi+\p_\mu (\frac{\p L}{\p\p_\mu \phi} \delta \phi) $. Thus, $\star \Theta=\frac{\p L}{\p\p_\mu \phi} \delta \phi$, and we recover well the definition \eqref{Noethercurrentclass}.}
\begin{equation} \label{eq53}
 J_\xi(g) = \Theta[\delta_{\xi} g,g] - i_\xi L\,. 
\end{equation}
The Noether current is on-shell closed, and thus locally there exists an $(n-2)$-form $Q_\xi (g)$ such that on-shell
\begin{equation} \label{eq54}
J_\xi(g)= -\extd Q_\xi (g),
\end{equation}
where $Q_\xi (g)$ is called the Noether charge in this formalism; it has nothing to do with \eqref{truenoether} and does not have to be confused with the conserved charge generating the action of the symmetry $\xi$ on the covariant phase space (denoted $H_\xi$). 
The symplectic current is defined as \cite{Lee:1990nz}
\begin{equation} \label{eq56}
\omega[\delta_1 g,\delta_2 g ;g]=\delta_1\Theta[\delta_2 g,g]-\delta_2\Theta[\delta_1 g,g],
\end{equation}
and the latter allows us to define the symplectic structure of the configuration space 
\begin{equation}
  \Omega [\delta_1 g, \delta_2 g; g]=\int_\Sigma \omega [\delta_1 g, \delta_2 g; g]\,,
\end{equation}
where $\Sigma$ is a Cauchy surface. 
The symplectic structure is exactly the infinitesimal\footnote{In the sense that it computes the charge difference between configurations $g$ and $g + \delta g $.} Hamiltonian generating the flow $g\rightarrow \delta _\xi g$
\begin{equation}
  \delta H_\xi =
  \Omega [\delta_\xi g,\delta g;g]\,.
\end{equation}
One can easily show \cite{Iyer:1994ys} that if $g$ satisfies the equations of motion
\begin{equation}
    \omega[\delta_\xi g,\delta g;g]=-\delta J_\xi +\extd(i_\xi \Theta ) \,,
\end{equation}
and moreover, if $\delta g$ satisfies the linearized equations of motion and $\xi$ is a symmetry, we also have
\begin{equation}
    \omega[\delta_\xi g,\delta g;g]=\extd k^{IW}_\xi (\delta g, g)=0 \,,
\end{equation}
with
\begin{equation}
     k^{IW}_\xi (\delta g, g) := i_\xi \Theta[\delta g,g] +\delta Q_\xi (g)\,.
\end{equation}
This formula gives a precise expression of the conserved $(n-2)$-form associated to $\xi$. The infinitesimal conserved charge is 
\begin{equation}
   \delta H_\xi= \int_{\partial \Sigma}  k_\xi (\delta g, g), 
\end{equation}
using Stoke's theorem and with $\p\Sigma$ being the boundary of the Cauchy surface $\Sigma$. The finite charge difference $H_\xi$ is obtained by an integral in configuration space.\\
Finally, the algebra of charges can be represented by a Dirac bracket as follows:
\begin{equation}  \label{intkgeneralexpression}
 \delta_\xi H_\zeta := \{H_\zeta, H_\xi\} = H_{[\zeta,\xi]} + \int_{\partial \Sigma}  k^{IW}_\zeta (\delta_\xi g, g).
\end{equation}
This algebra of charges is valid only when the charges are integrable. The second term on the right-hand side is recognized as a central extension, which cannot be absorbed in a redefinition of the generators and thus has important physical consequences.\\
This formalism contains ambiguities in the definition of the conserved $(n-2)$-form coming from the arbitrariness in the symplectic potential $\Theta \rightarrow \Theta + \extd Y$. Other physical arguments can be used to fix these ambiguities. For example, the requirement to have a well defined variational principle. \\
Another prescription was proposed by Barnich, Brandt and Comp{\`e}re (BBC) \cite{Barnich:2001jy, Barnich:2007bf} which we will elaborate on a bit now. 
Under gauge transformations, one can write, using Bianchi identities,
\begin{equation} \label{eq51}
 E(g) \delta_\xi g = \extd S_\xi(E(g),g),
\end{equation}
where $S$ is the weakly vanishing Noether current. 
The latter is conserved and vanishes on-shell.
For example, for a pure gravity theory, one has $E(g) \delta_\xi g = \nabla_\mu (2 \xi_\nu E^{\mu \nu}) - 2 \xi_\nu \nabla_\mu E^{\mu \nu}$, where the last term vanishes by virtue of Bianchi identities, and hence in that case 
$S_\xi^\mu = 2 \xi_\nu E^{\mu \nu}$. 
One can act on the weakly vanishing Noether current with a contracting homotopy operator, yielding an $(n-2)$-form denoted $k^{BB}_\xi (\delta g, g)$ which is conserved for $\xi$ being a reducibility parameter, $g$ a solution of the equations of motion, $\delta g$ a solution of the linearized equations of motion. 
In essence, this operator is the inverse of the exterior derivative $\extd$ (see e.g. \cite{Chen:2013aza} for an explicit expression).
One advantage of this procedure is that it provides a definition of charges depending only on the equations of motion
of the Lagrangian, and not on boundary terms.\\
The two formalisms are related by\footnote{The E-term is not the equations of motion in this last paragraph.}  
\begin{equation} \label{eq59}
  k^{BB}_\xi (\delta g, g) = k^{IW}_\xi (\delta g, g) + E(\delta_\xi g, \delta g),
\end{equation}
in which the expression of $E(\delta_\xi g, \delta g)$ is known explicitly (see e.g. (3.7) of \cite{Azeyanagi:2009wf}).
Please note that this ambiguity is not relevant for exact symmetries, having $\delta_\xi g = 0$, but may yield distinct results in the asymptotic context (see \cite{Azeyanagi:2009wf} for one such example in Kerr/CFT).\\
These last considerations end the chapter on the canonical analysis of charges. In the next section, we illustrate these points on the example of asymptotically flat spacetimes in 2+1 dimensions. 

\section{Flat Space Holography}

This section of the lecture notes is devoted to a putative holographic principle governing asymptotically flat spacetimes. There has been quite a substantial amount of research going on for the last couple of years, especially in three-dimensional Einstein gravity, indicating that there is, indeed, a holographic principle in flat space\footnote{For one of first papers making this idea manifest in three spacetime dimensions see \cite{Bagchi:2010eg}.}. Since these checks require the knowledge of the basic symmetries of a putative dual quantum field theory the first step usually consists in finding appropriate boundary conditions that yield interesting boundary dynamics\footnote{One can also formulate interesting boundary conditions in higher-derivative theories of gravity in flat space such as e.g. in \cite{Setare:2017mry}.} such as e.g. \cite{Barnich:2006av,Afshar:2015wjm,Afshar:2016kjj,Detournay:2016sfv,Ammon:2017vwt,Grumiller:2017sjh}. One can then use these symmetries to perform non-trivial checks of a possible holographic correspondence.
These checks include for example a derivation and matching of a Cardy-like formula (including logarithmic corrections) of cosmological solutions in flat space\footnote{These cosmological solutions in asymptotically flat spacetimes were first described in \cite{Cornalba:2002fi,Cornalba:2003kd}.} \cite{Barnich:2012xq, Bagchi:2012xr,Bagchi:2013qva,Riegler:2014bia,Fareghbal:2014qga,Basu:2017aqn,Fuentealba:2017fck}, calculating holographic entanglement entropy \cite{Bagchi:2014iea,Basu:2015evh,Jiang:2017ecm} including logarithmic corrections \cite{Fareghbal:2017ujy}, one-loop (higher-spin) partition functions in flat space \cite{Barnich:2015mui,Campoleoni:2015qrh,Campoleoni:2016vsh} or the computation of holographic stress tensor correlation functions \cite{Bagchi:2015wna,Asadi:2016plj}.\\
In the following we want to review some aspects of such holographic principle and in particular we want to present how a canonical analysis can help identifying a putative dual quantum field theory for asymptotically flat spacetimes  in three dimensions. We will do so by using both the metric as well as the Chern-Simons formulation.

\subsection{The Metric Perspective}

In gravitation, to describe a phase space it is necessary to characterize the dynamics and specify the fall-off behavior of fields, i.e. boundary conditions. The latter will ensure that the content of the phase space is physically consistent.\\
Usually, well defined boundary conditions require to: contain physically interesting spacetimes, have non-trivial asymptotic symmetries, and lead to finite and integrable charges. It is usually very difficult to find such a set. For pedagogical reasons, we will present one of those set in the metric formulation and verify that it satisfies, indeed, these requirements. \\
We choose to consider the $\mathfrak{bms}_3$ boundary conditions arising in the descriptions of asymptotically flat spacetimes in 2+1 dimensions. 

\subsubsection{$\mathfrak{bms}_3$ boundary conditions}

The so called $\mathfrak{bms}_3$ boundary conditions (BCs) were proposed in \cite{Barnich:2006av}, see also \cite{Bagchi:2012yk},
\begin{subequations}\label{BC}
\begin{align}
& g_{uu}=h_{uu}+O\left(r^{-1}\right)\,, && g_{ur}=-1 + \frac{h_{ur}}r+O\left(r^{-2}\right)\,,\\
&  g_{u\phi}=h_{u\phi}+ O\left(r^{-1}\right)\,,&& g_{rr}= \frac{h_{rr}}{r^2}+O\left(r^{-3}\right)\,,\\
&  g_{r\phi}=h_1(\phi) + \frac{h_{r\phi}}r+O\left(r^{-2}\right)\,,&&  g_{\phi\phi}=r^2 +(h_2(\phi)+u\,h_3(\phi))r+O\left(1\right),
\end{align}
\end{subequations}
where $r$ is the radial coordinate and $h_{\mu\nu}$ are functions of the retarded time $u$ and the angle $\phi$ (taken $2\pi$-periodic). Minkowski spacetime, written in Eddington-Finkelstein coordinates, 
\begin{equation}
\extd s^2=-\extd u^2-2\extd u\,\extd r+r^2\,\extd\phi,
\end{equation}
is included in these $\mathfrak{bms}_3$ BCs, obtained by taking $h_{uu}=-1$ and setting all the other arbitrary functions to zero. 
Also, another set of exact solutions of Einstein equations \cite{Barnich:2012aw} is included
\begin{equation}\label{SolExact}
\extd s^2=\mathcal{M}(\phi)\,\extd u^2-2\, \extd u\, \extd r+2\left( \frac{\mathcal{M}'(\phi)}2\, u + \mathcal{N}(\phi)\right) \, \extd u\,\extd\phi +r^2\,\extd\phi^2\,.
\end{equation}
These solutions are the analog of the Ba{\~n}ados solutions obtained in locally AdS$_3$ spacetimes \cite{Banados:1998gg}. In this family, the flat space cosmology \cite{Cornalba:2002fi} is included by taking $\mathcal{M}(\phi)=8M$, $\mathcal{N}(\phi)=4J$ with $M>0$ and $J\neq 0$,  
\begin{equation}
\extd s^2 = -2 \extd r \extd u+8 M \extd u^2+8 J \extd u \extd\phi+r^2 \extd\phi^2\,.
\end{equation}
It possesses a cosmological horizon located at $r_c=\sqrt{2J^2/M}$. 
For more information on flat space cosmology, see for example \cite{Barnich:2012aw,Cornalba:2002fi,Cornalba:2003kd}. 
So, these BCs satisfy the first requirement, namely including known interesting solutions. 
In the following, we determine the asymptotic symmetry algebra. 
 
\subsubsection{Asymptotic symmetries}

The first mathematical objects relevant for this question are the asymptotic Killing vectors (AKVs). 
As an example, we compute them explicitly for the exact solutions \eqref{SolExact} and then present them for the $\mathfrak{bms}_3$ BCs \eqref{BC}.\\
First, let us recall that a Killing vector $\xi$ satisfies
\begin{equation}
\mathsterling_{\xi} g=0\,.
\end{equation}
The Lie derivative is in components given by $\mathsterling_\xi g_{\mu\nu}=\nabla_\mu\xi_\nu+\nabla_\nu\xi_\mu$, or equivalently  $\mathsterling_\xi g_{\mu\nu}=\xi^\rho\p_\rho g_{\mu\nu}+g_{\mu\rho}\p_\nu \xi^\rho+g_{\nu\rho}\p_\mu \xi^\rho$.\\
The asymptotic Killing vectors (AKVs) only preserve the structure of the spacetime asymptotically, i.e. they leave the asymptotic form of the metric invariant up to a redinition of its arbitrary functions. \\
So, in our working example \eqref{SolExact}, the AKV's action will leave in metric untouched up to a redefinition of the functions $\mathcal{M},\,\mathcal{N}$.
We start with the most general candidate for the AKVs 
\begin{equation}
\xi=\xi^u(u,r,\phi )\p_u+\xi^r(u,r,\phi )\p_r+\xi^\phi(u,r,\phi )\p_\phi\,. 
\end{equation} 
First, we impose 
\begin{equation}
\mathsterling_\xi g _{rr}=0\,,\qquad \mathsterling_\xi g _{r\phi}=0\,,\qquad \mathsterling_\xi g _{\phi\phi }=0\,.
\end{equation}
This leads to the following constraints
\begin{subequations}
\begin{align}
& \mathsterling_\xi g_{rr}= -2\p_r \xi^u \Rightarrow \xi^u(u,r,\phi)=F(u,\phi),\\
&  \mathsterling_\xi g_{r\phi}= -\p_\phi F(u,\phi)+r^2 \p_r\xi^\phi(u,r,\phi) \Rightarrow \xi^\phi(u,r,\phi)=G(u,\phi)-\frac{\p_\phi F(u,\phi)}{r},\\ \nonumber
& \mathsterling_\xi g_{\phi\phi}=2\left(r \, \xi^r(u,r,\phi)+g_{u\phi} \p_\phi F(u,\phi)+r^2 \, \p_\phi\xi^\phi(u,r,\phi)
\right) \\
& \qquad \qquad \qquad 
\Rightarrow \xi^r(u,r,\phi)=-r \, \p_\phi\xi^\phi(u,r,\phi)-\frac1rg_{u\phi} \p_\phi F(u,\phi)\,.
\end{align}
\end{subequations}
We impose now the last conditions. Namely, 
\begin{equation}
\mathsterling_\xi g _{ur}=0\,,\qquad
\mathsterling_\xi g _{uu }=\tilde M(\phi)
\,,\qquad 
\mathsterling_\xi g _{u\phi}=\frac{\tilde M(\phi)}2+\tilde N(\phi)\,,
\end{equation}
where $\tilde M(\phi),\tilde N(\phi)$ are arbitrary functions of $\phi$.
As all the $r$-dependences of the AKV's components are fixed, we can look order by order in $r$. 
The $r^2$-component of $\mathsterling_\xi g _{u\phi}$ gives
\begin{equation*}
\p_u G(u,\phi)=0 \Rightarrow G(u,\phi)=g(\phi)\,. 
\end{equation*}
After this, we impose 
\begin{align*}
& \mathsterling_\xi g_{ur}=\p_\phi g(\phi)- \p_uF(u,\phi) \Rightarrow F(u,\phi)=f(\phi)+u\,\p_\phi g(\phi)\,.
\end{align*}
The rest of the constraints are satisfied. Finally, the AKVs for the family of solutions \eqref{SolExact} take the form
\begin{equation}\label{AKVSolExact}
\xi=F(u,\phi)\p_u+ \left( -\p_\phi g(\phi)
\,r+\p_\phi^2 F(u,\phi)- \frac{g_{u\phi}\p_\phi F(u,\phi)}r
\right)\p_r+
\left(g(\phi)-\frac{\p_\phi F(u,\phi)}{r}\right)\p_\phi,
\end{equation}
with $F(u,\phi)=f(\phi)+u\,\p_\phi g(\phi)$. These AKVs depend on two arbitray functions $f(\phi),g(\phi)$.\\
Now, we consider the AKVs for the $\mathfrak{bms}_3$ BCs \eqref{BC}. The strategy to derive them is exactly the same as the one presented above. Actually, their leading terms are exactly the same as in \eqref{AKVSolExact}. The precise subleadings can be computed acting on the $\mathfrak{bms}_3$ BCs. We present the AKVs of $\mathfrak{bms}_3$ BCs \eqref{BC} using the periodicities of the functions in $\phi$ to decompose them in Fourier modes. We have\footnote{The two independent functions $f(\phi)$ and $g(\phi)$ of \eqref{AKVSolExact} are developed in two disctinct sets of modes.}
\begin{subequations}\label{AKVBMS}
\begin{align}
& \ell_n= \, e^{i\,n\,\phi}\left[\left(i\,n\,u +O\left(r^{-1} \right)\right)\p_u+\left(-i\,n \,r+O\left(1 \right)\right)\p_r+\left(1+O\left(r^{-1} \right)\right)\p_\phi\right], \\
&m_n= \left(  e^{i\,n\,\phi}+O\left(r^{-1} \right) \right)\p_u + O\left(1 \right) \p_u+ O\left(r^{-1} \right) \p_\phi\,.
\end{align}
\end{subequations}
We note that the six Killing vectors of Minkowski spacetime are recovered with $n=\pm1,0$ and form an $\mathfrak{isl}(2,\mathbb{R})$ algebra.\\
Now, we consider the algebra satisfied by the AKVs. 
In a well defined set of BCs, the Lie bracket of the AKVs, or the modified one when the AKVs are state-dependant \cite{Barnich:2010eb} (or ``adjusted'' \cite{Compere:2015knw}), have to constitute an algebra, i.e. have to close and satisfy the Jacobi identities. The commutation relations are straightforwardly computed \footnote{Convention: $[\xi_1,\xi_2 ]=\mathsterling_{\xi_1}\xi_2$.}
\begin{subequations}
\begin{align}
& i [\ell_n,\ell_m ]= (n-m)\ell_{n+m},\\
& i [\ell_n,m_m ]= (n-m)m_{n+m},\\
& i [m_n,m_m ]= 0\,.
\end{align}
\end{subequations}
It is indeed closed and one can easily check that Jacobi identities are satisfied. We have thus a well defined algebra composed of a de Witt algebra, generated by $\ell_n$'s, and an abelian ideal, called the supertranslations, generated by $m_n$'s.  
This algebra is precisely the $\mathfrak{bms}_3$ algebra.\\
This is not yet the asymptotic algebra of symmetries (ASA). 
The latter consists in the allowed diffeomorphisms modulo the trivial diffeomorphisms. We've already found the allowed diffeomorphisms for our BCs, i.e. the AKVs. The trivial diffeomorphisms demand a little more work. 
The AKVs are asymptotically the reducibility parameters and thus, there are charges associated to them. 
A trivial diffeomorphism has its associated charge equals to zero. So we compute the charges to determine the trivial AKVs.  \\
In order to compute the charges, we must specify the dynamics of our system - for example Einstein-Hilbert theory.
In this case, the charges can be computed for example using the BBC method \footnote{Geoffrey Comp{\`e}re has encoded the computations of charges in some theories. The package can be found at  \url{http://www.ulb.ac.be/sciences/ptm/pmif/gcompere/package.html}}, 
\begin{align}
\delta\Lt_n:=\delta H_{\ell_n} & =\frac{-1}{32\pi\,G}\int_0^{2\pi}d\phi \,e^{i\,n\,\phi}\delta \big( 2 i n h_1'(\phi )+2 h_1(\phi ) h_3(\phi )+n^2 h_1(\phi )+2 i n h_2 (\phi )-4 h_{u\phi}(\phi )\big),\\
\delta\Mt_n:=\delta H_{m_n}& = \frac{1}{16\pi\,G}\int_0^{2\pi}d\phi\, e^{i\,n\,\phi}\delta (h_3(\phi)+h_{uu}(\phi) ),
\end{align}
where we have imposed the equations of motion, giving $h_{ur}(u,\phi)=-\frac12\p_u h_{rr}(u,\phi)$, $h_{uu}(u,\phi)=h_{uu}(\phi)$ and $h_{u\phi}(u,\phi)=h_{u\phi}(\phi)-\frac{1}{4} \p_u\p_\phi h_{rr}(u,\phi )-\frac{1}{2} \p_u h_{r\phi}(u,\phi )+\frac{1}{2} u h_{uu}'(\phi )$. The charges are indeed finite and integrable. 
To go back to the trivial diffeomorphisms, it turns out that they are hidden in some of the subleading terms in the AKVs (not explicitly written in \eqref{AKVBMS}).\\
The last task we have to face is the computation of the putative central charges. 
They appear only when we consider the commutation relations of the charges associated to the AKVs and not only the AKVs themselves \eqref{intkgeneralexpression}. In the case of our example of $\mathfrak{bms}_3$ boundary conditions in Einstein-Hilbert gravity, the asymptotic symmetry algebra of the charges is
\begin{subequations}
\begin{align}
& i [\Lt_n,\Lt_m ]= (n-m)\Lt_{n+m}\,,\\
& i [\Lt_n,\Mt_m ]= (n-m)\Mt_{n+m}+\frac1{4G}(n^3-n)\delta_{n+m,0}\,,\\
& i [\Mt_n,\Mt_m ]= 0\,,
\end{align}
\end{subequations}
with $G$ the Newton constant. \\
This algebra of symmetries is the $\mathfrak{bms}_3$ algebra. This algebra is isomorphic to the 2D Galilean Conformal Algebra ($\mathfrak{gca}_2$). Because of this isomorphism a new duality was proposed in \cite{Bagchi:2010eg} involving the $\mathfrak{bms}_3$ and $\mathfrak{gca}_2$ algebra in the same spirit as the $AdS_3/CFT_2$ correspondence. Soon after this conjecture there were a couple of explicit checks supporting it. Maybe the most well known is the matching of the entropy of the cosmological horizon of flat space cosmologies via a Cardy-like formula derived using $\mathfrak{gca}_2$ methods \cite{Barnich:2012xq,Bagchi:2012xr}.

\subsection{The Chern-Simons Perspective}

Let us now focus on the Chern-Simons perspective of the things considered previously in this section of the lecture notes. The goal of this part of the lecture notes is to explain how to describe three-dimensional Einstein gravity with vanishing cosmological constant using the Chern-Simons formulation and define boundary conditions whose corresponding canonical boundary charges yield the three-dimensional Bondi-Metzner-Sachs ($\mathfrak{bms}_3$) algebra \cite{Bondi:1962px,Sachs:1962zza}. \\
As already mentioned in Section~\ref{ch:ChernSimonsFormulationofGravity} Einstein gravity with vanishing cosmological constant in three dimensions is described in terms of a Chern-Simons theory where the gauge field $\mathcal{A}$ takes values in $\mathfrak{isl}(2,\mathbb{R})$. The Chern-Simons level $k$ is related to Newton's constant $G$ in three dimensions via $k=\frac{1}{4 G}$.
Using a basis for $\mathfrak{isl}(2,\mathbb{R})$ with generators $\Lt_n$ and $\Mt_n$ with $n=0,\pm1$ that have the following non-vanishing Lie brackets:
	\begin{subequations}\label{eq:isl2RBasis}
	\begin{align}
		[\Lt_n,\Lt_m]&=(n-m)\Lt_{n+m},\\
		[\Lt_n,\Mt_n]&=(n-m)\Mt_{n+m}.
	\end{align}
	\end{subequations}
The corresponding invariant bilinear form appearing in \eqref{eq:ChernSimonsAction} is given by $\langle \Lt_n\Lt_m\rangle=\langle \Mt_n\Mt_m\rangle=0$ as well as
	\begin{subequations}\label{eq:ISLInvBilForm}
	\begin{align}
		\langle \Lt_n\Mt_m\rangle & =-2\left(
			\begin{array}{c|ccc}
				  &\Mt_1&\Mt_0&\Mt_{-1}\\
				\hline
				\Lt_1&0&0&1\\
				\Lt_0&0&-\frac{1}{2}&0\\
				\Lt_{-1}&1&0&0
			\end{array}\right).
	\end{align}		
	\end{subequations}
Let us assume that the topology of the manifold is that of a solid cylinder. In addition we choose coordinates such that there is a radial direction $0\leq r<\infty$ and the boundary of the cylinder is parametrized by a retarded time coordinate $-\infty<u<\infty$ as well as an angular coordinate $\varphi\sim\varphi+2\pi$.\\
We then fix the radial dependence of gauge fields $\mathcal{A}$ as
	\begin{equation}\label{eq:RadialDep}
		\mathcal{A}(r,u,\varphi)=b^{-1}(r)\left[a(u,\varphi)+\extd\,\right]b(r),	
	\end{equation}
with
	\begin{equation}
		a(u,\varphi)=a_\varphi(u,\varphi)\extd\varphi+a_u(u,\varphi)\extd u.
	\end{equation}
This gauge fixing has the advantage that the $F_{\rho i}$ components of the EOM, for $i=t,\varphi$ are automatically satisfied. This can be seen by a short direct calculation. Imposing the gauge choice \eqref{eq:RadialDep} the equations of motion $F=0$ can be written as
    \begin{equation}
        F_{\rho i} = \partial_\rho(b^{-1}a_ib)-\underbrace{\partial_i(b^{-1}\extd b)}_{0}+[b^{-1}\extd b,b^{-1}a_ib].
    \end{equation}
Explicitly evaluating the partial derivative as well as the commutator one finds that
    \begin{equation}
        F_{\rho i} = \partial_\rho b^{-1}a_ib+b^{-1}\partial_\rho bb^{-1}a_ib=\partial_\rho b^{-1}a_ib+\partial_\rho(\underbrace{b^{-1}b}_{\unity})b^{-1}a_ib-\partial_\rho b^{-1}\underbrace{bb^{-1}}_{\unity}a_ib=0.
    \end{equation}
Similarly the EOM for $F_{ij}$ simplify to
    \begin{equation}
        F_{ij}=b^{-1}\left(\partial_i a_j-\partial_j a_i+[a_i,a_j]\right)b=0,
    \end{equation}
so that effectively one only has to satisfy $\partial_i a_j-\partial_j a_i+[a_i,a_j]=0$.\\
It is important to note that different choices of the group element $b$ will yield different geometrical interpretations. One very popular choice of $b$ is given by e.g.
	\begin{equation}
		b(r)=e^{\frac{r}{2}\Mt_{-1}}.
	\end{equation}
Using this gauge one can relate the boundary conditions\footnote{These boundary conditions will be specified in the next subsection.} \eqref{eq:NewFSBCs}, to the boundary conditions used in the previous section in the metric formalism. This is done by extracting the dreibein $e$ from the Chern-Simons connection $\mathcal{A}$ via
	\begin{equation}
		\mathcal{A}=\omega^aL_a+e^aM_a,
	\end{equation}
for $a=0,\pm1$. Then using
	\begin{equation}
		\eta_{ab}=-2\left(
			\begin{array}{c|ccc}
				  &M_1&M_0&M_{-1}\\
				\hline
				M_1&0&0&1\\
				M_0&0&-\frac{1}{2}&0\\
				M_{-1}&1&0&0
			\end{array}\right),
	\end{equation}
one can recover the metric formulation via 
	\begin{equation}
		g_{\mu\nu}=\eta_{ab}e^a_\mu e^b_\nu.
	\end{equation}
For the boundary conditions \eqref{eq:NewFSBCs} this leads to the following metric:
	\begin{equation}\label{eq:AdSBMSMetricNew}
		\extd s^2=\tilde{\mathcal{M}}\extd u^2+2\tilde{\mathcal{N}}\extd u\extd\varphi-2\extd r\extd u+r^2\extd\varphi^2,
	\end{equation}
where in terms of the variables used in \eqref{eq:NewFSBCs} one has
    \begin{equation}
        \tilde{\mathcal{M}}=-\frac{\pi}{k}\mathcal{M},\qquad\tilde{\mathcal{N}}=-\frac{\pi}{k}\mathcal{N}.
    \end{equation}
    
\subsubsection{Boundary Conditions}

After having completely specified our specific setup we are now ready to formulate boundary conditions. One can write down boundary conditions in terms of the gauge field $a$ as\footnote{One might wonder why we chose to define the functions $\mathcal{M}$ and $\mathcal{N}$ including these prefactors of $-\frac{\pi}{k}$. This is done with a bit of hindsight already as including such a prefactor allows one to very efficiently determine the Dirac bracket algebra of the canonical boundary charges.}
	\begin{subequations}\label{eq:NewFSBCs}
	\begin{align}
		a_\varphi&=\Lt_1-\frac{\pi}{k}\mathcal{M}\Lt_{-1}-\frac{\pi}{k}\mathcal{N}\Mt_{-1},\\
		a_u&=\Mt_1-\frac{\pi}{k}\mathcal{M}\Mt_{-1}.
	\end{align}
	\end{subequations}
where the functions $\mathcal{M}$ and $\mathcal{N}$ are arbitrary functions of $u$ and $\varphi$.\\
Looking at the equations of motion $F=0$ one obtains very simple constraints on the (retarded) time evolution of the functions $\mathcal{M}$ and $\mathcal{N}$ as
	\begin{equation}\label{eq:EOM}
		\partial_u\mathcal{M}=0,\qquad\partial_u\mathcal{N}=\partial_\varphi\mathcal{M}.
	\end{equation}
That means that on-shell these functions can be written as
    \begin{equation}\label{eq:StateOnShell}
    \mathcal{M} =  \mathcal{M}(\varphi),\qquad \mathcal{N} = \mathcal{L}(\varphi) +u\mathcal{M}'.
    \end{equation}
After having properly specified a set of boundary conditions the next step is to determine the set of gauge transformations that preserves these boundary conditions. This is most easily done by making an ansatz of the form
	\begin{equation}\label{eq:GaugeParameter}
		\bar{\epsilon}(r,u,\varphi)=b^{-1}\left[\sum\limits_{a=-1}^1\epsilon^a(u,\varphi)\Lt_a+\sigma^a(u,\varphi)\Mt_a\right]b.
	\end{equation}
In terms of this ansatz the gauge transformations (including proper and non-trivial ones) that preserve the boundary conditions \eqref{eq:NewFSBCs} are given by
	\begin{subequations}\label{eq:BCPGTs}
	\begin{align}
		\epsilon^1&=\epsilon,& \epsilon^0&=-\epsilon',&
		\epsilon^{-1}&=-\frac{\pi}{k}\mathcal{M}\epsilon+\frac{\epsilon''}{2},\\
		\sigma^1&=\sigma,&\sigma^0&=-\sigma',&
		\sigma^{-1}&=-\frac{\pi}{k}\mathcal{N}\epsilon-\frac{\pi}{k}\mathcal{M}\sigma+\frac{\sigma''}{2},
	\end{align}
	\end{subequations}
where the functions $\epsilon$ and $\sigma$ depend on $u$ and $\varphi$ and a prime denotes differentiation with respect to $\varphi$. In addition these functions have to satisfy
	\begin{equation}\label{eq:GaugeParametersTimeEvolution}
		\partial_u\epsilon=0,\qquad\partial_u\sigma=\partial_\varphi\epsilon.
	\end{equation}
That means that these gauge parameters can also be written as
    \begin{equation}
    \epsilon =  \epsilon(\varphi),\qquad\sigma = \sigma(\varphi) +u\epsilon'.
    \end{equation} 
The fields $\mathcal{M}$ and $\mathcal{L}$ then transform under the gauge transformations \eqref{eq:BCPGTs} as
	\begin{subequations}\label{eq:GaugeTrafos}
	\begin{align}
		\delta_{\bar\epsilon}\mathcal{M}&=\epsilon\mathcal{M}'+2\mathcal{M}\epsilon'-\frac{k}{2\pi}\epsilon''',\\
		\delta_{\bar\epsilon}\mathcal{L}&=\sigma\mathcal{M}'+2\mathcal{M}\sigma'-\frac{k}{2\pi}\sigma'''+\epsilon\mathcal{L}'+2\mathcal{L}\epsilon'.
	\end{align}
	\end{subequations}
The corresponding canonical boundary charges are obtained by functionally integrating \cite{Henneaux:1992,Blagojevic:2002aa}
    \begin{equation}
        \delta Q[\bar{\epsilon}]=\frac{k}{2\pi}\int\extd\varphi\,\left\langle\epsilon \,\delta\mathcal{A}_\varphi\right\rangle.
    \end{equation}
For the boundary conditions \eqref{eq:NewFSBCs} one obtains 
    \begin{equation}\label{eq:CanChargesVariation}
        \delta Q[{\bar\epsilon}]=\int\extd\varphi\left(\epsilon\delta\mathcal{L}+\sigma\delta\mathcal{M}\right),
    \end{equation}
which can be functionally integrated to yield 
    \begin{equation}\label{eq:CanChargesEinstein}
        Q[{\bar\epsilon}]=\int\extd\varphi\left(\epsilon\mathcal{L}+\sigma\mathcal{M}\right).
    \end{equation}
    
\subsubsection{Asymptotic Symmetries}\label{sec:FSASA}

Having the canonical boundary charges in a form like \eqref{eq:CanChargesEinstein} is very beneficial when it comes to determining the Dirac brackets of these canonical charges. Reason being that such a form allows one to directly read off the Dirac bracket algebra of the functions $\mathcal{M}$ and $\mathcal{L}$ from \eqref{eq:GaugeTrafos} using e.g. 
	\begin{equation}
		\{\mathcal{L}(\varphi),\mathcal{M}(\bar{\varphi})\}_{\textrm{D.B.}}=-\delta_{\epsilon}\mathcal{M}(\bar{\varphi})
		\Bigr|_{\partial_{\bar{\varphi}}^n\epsilon(\bar{\varphi})=(-1)^n\partial_\varphi^n\delta(\varphi-\bar{\varphi})}.
	\end{equation}
This trick works because the infinitesimal transformations \eqref{eq:GaugeTrafos} are related to the Dirac brackets of the canonical boundary charges as 
    \begin{equation}
        -\delta_{\epsilon}\mathcal{M}(\bar{\varphi})=\{Q[\epsilon],\mathcal{M}(\bar{\varphi})\}_{\textrm{D.B.}},
    \end{equation}
which reduces to
    \begin{equation}
        -\delta_{\epsilon}\mathcal{M}(\bar{\varphi})=
		\int\extd\varphi\,\epsilon(\varphi)\{\mathcal{L}(\varphi),\mathcal{M}(\bar{\varphi})\}_{\textrm{D.B.}},
	\end{equation}
in case all coefficients in front of the canonical boundary charges are equal to one.\\  
After employing this trick one finds the following non-vanishing Dirac brackets for the state dependent functions:
	\begin{subequations}
	\begin{align}
	\{\mathcal{L}(\varphi),\mathcal{L}(\bar{\varphi})\}_{\textrm{D.B.}}&=2\mathcal{L}\delta'-\delta\mathcal{L}',\\
	\{\mathcal{L}(\varphi),\mathcal{M}(\bar{\varphi})\}_{\textrm{D.B.}}&=2\mathcal{M}\delta'-\delta\mathcal{M}'-\frac{k}{2\pi}\delta''',
	\end{align}
	\end{subequations}
where all functions appearing on the r.h.s are functions of $\bar{\varphi}$ and prime denotes differentiation with respect to the corresponding argument. Moreover ${\delta\equiv\delta(\varphi-\bar{\varphi})}$ and $\delta'\equiv\partial_\varphi\delta(\varphi-\bar{\varphi})$. Expanding the fields and delta distribution in terms of Fourier modes as
	\begin{subequations}\label{eq:EinsteinFourierModes}
	\begin{align}
		\mathcal{M}&=\frac{1}{2\pi}\sum\limits_{n\in\mathbb{Z}}\left(\Mt_n-\frac{k}{2}\delta_{n,0}\right)e^{-in\varphi},&
		\mathcal{L}&=\frac{1}{2\pi}\sum\limits_{n\in\mathbb{Z}}\Lt_ne^{-in\varphi},\\
		\delta&=\frac{1}{2\pi}\sum\limits_{n\in\mathbb{Z}}e^{-in(\varphi-\bar{\varphi})},
	\end{align}
	\end{subequations}
and then replacing the Dirac brackets with commutators using $i\{\cdot,\cdot\}_{\textrm{D.B.}}\rightarrow[\cdot,\cdot]$ one obtains the following non-vanishing commutation relations:
	\begin{subequations}\label{eq:ASAPrelim}
	\begin{align}
		[\Lt_n,\Lt_m]&=(n-m)\Lt_{n+m},\\
		[\Lt_n,\Mt_m]&=(n-m)\Mt_{n+m}+\frac{c_M}{12}n(n^2-1)\delta_{n+m,0},
	\end{align}
	\end{subequations}
with $c_M=12k$ which is exactly the $\mathfrak{bms}_3$ algebra with the central extension first found in \cite{Barnich:2006av} using the metric formulation.

\section{Most General Flat Space boundary Conditions}

In order to get a little bit more accommodated with imposing boundary conditions as well as the physical consequences that follow from imposing different boundary conditions we want to present another set of boundary conditions. This set is in a sense the most general set of boundary conditions that one can impose in a Chern-Simons formulation in asymptotically flat spacetimes\footnote{There is also a corresponding set of boundary conditions for asymptotically AdS$_3$ spacetimes \cite{Grumiller:2016pqb}.} \cite{Grumiller:2017sjh}.

\subsection{The Boundary Conditions}

The setup will be exactly the same as in the previous section. The only thing that we will be changing are the boundary conditions of the $\mathfrak{isl}(2,\mathbb{R})$ valued gauge field $\mathcal{A}$. We again choose a gauge as \eqref{eq:RadialDep}, however, in contrast to the previous section we will not give an explicit expression of $b(r)$ this time as we will only be interested in the canonical charges and the resulting asymptotic symmetry algebra from a Chern-Simons perspective, where $b(r)$ does not really have any influence. Reason being that $b(r)$ appears neither in the boundary condition preserving gauge transformations nor the expressions for the canonical boundary charges. Roughly speaking, the exact form of the group element $b(r)$ is only relevant if one is interested in a metric formulation of the asymptotic symmetries.\\
Now without further ado let us present the boundary conditions originally first presented in \cite{Grumiller:2017sjh}
\begin{align}\label{abcs}
a_{\varphi} & = - \left(\mathcal{M}^+ L_+ - 2\mathcal{M}^0 L_0 + \mathcal{M}^- L_- + \mathcal{L}^+ M_+ - 2 \mathcal{L}^0 M_0 + \mathcal{L}^- M_- \right) \\
a_u & = \mu_L^n L_n + \mu_M^n M_n.
\label{eq:bms6}
\end{align}
The six functions $\mathcal{M}^a$ and $\mathcal{L}^a$ are in a holographic context usually related to vacuum expectation values of operators in the dual quantum field theory and the functions $\mu_L^n$ and $\mu_M^n$ are interpreted as chemical potentials/sources. We assume all functions to be arbitrary functions of the boundary coordinates and demand in additional that the chemical potentials are fixed, i.e. $\delta\mu_L^n=\delta\mu_M^n=0$.\\
This connection solves the flatness condition $\extd A + A \wedge A = 0$ if the functions $\mathcal{M}^a$ and $\mathcal{L}^a$ satisfy
\begin{subequations}
\label{eq:bms5}
\begin{align}
\partial_u \mathcal{L}^{\pm} & = \pm \mu_L^0 \mathcal{L}^{\pm} \pm 2 \mu_L^{\pm} \mathcal{L}^0 \pm \mu_M^0 \mathcal{M}^{\pm} \pm 2 \mu_M^{\pm} \mathcal{M}^0 - \partial_\varphi \mu_M^{\pm} \\
\partial_u \mathcal{L}^{0} & = \mu_L^+ \mathcal{L}^{-} - \mu_L^{-} \mathcal{L}^+ + \mu_M^+ \mathcal{M}^{-} - \mu_M^{-} \mathcal{M}^+ + \frac{1}{2}\partial_\varphi \mu_M^{0} \\
\partial_u \mathcal{M}^{\pm} & = \pm \mu_L^0 \mathcal{M}^{\pm} \pm 2 \mu_L^{\pm} \mathcal{M}^0 - \partial_\varphi \mu_L^{\pm} \\
\partial_u \mathcal{M}^{0} & = \mu_L^+ \mathcal{M}^{-} - \mu_L^{-} \mathcal{M}^+  + \frac12 \partial_\varphi \mu_L^{0} \,.
\end{align}
\end{subequations}
The procedure to determine the asymptotic symmetries associated to these boundary conditions is exactly the same as in the previous sections. Thus the next steps are determining the boundary conditions preserving gauge transformations and consequently also the canonical boundary charges. In order to determine the boundary condition preserving gauge transformations one makes again the ansatz
\begin{equation}
\epsilon = b^{-1} (\epsilon_M^n M_n + \epsilon_L^n L_n) b\,.
\end{equation}
Since the $a_\varphi$-component of the connection and it contains the most general form possible, there are no restrictions on the $\varphi$-dependence of the gauge parameters. Thus the state-dependent functions transform as
\begin{subequations}\label{trafos}
\begin{align}
\delta \mathcal{L}^{\pm} & = \pm \epsilon_L^0 \mathcal{L}^{\pm} \pm 2 \epsilon_L^{\pm} \mathcal{L}^0 \pm \epsilon_M^0 \mathcal{M}^{\pm} \pm 2 \epsilon_M^{\pm} \mathcal{M}^0 - \partial_\varphi \epsilon_M^{\pm} \\
\delta \mathcal{L}^{0} & = \epsilon_L^+ \mathcal{L}^{-} - \epsilon_L^{-} \mathcal{L}^+ + \epsilon_M^+ \mathcal{M}^{-} - \epsilon_M^{-} \mathcal{M}^+ + \frac{1}{2}\partial_\varphi \epsilon_M^{0} \\
\delta \mathcal{M}^{\pm} & = \pm \epsilon_L^0 \mathcal{M}^{\pm} \pm 2 \epsilon_L^{\pm} \mathcal{M}^0 - \partial_\varphi \epsilon_L^{\pm} \\
\delta \mathcal{M}^{0} & = \epsilon_L^+ \mathcal{M}^{-} - \epsilon_L^{-} \mathcal{M}^+  + \frac12 \partial_\varphi \epsilon_L^{0} \,.
\end{align}
\end{subequations}
The $a_u$-component of the connection according to our boundary conditions has to satisfy $\delta a_u=0$.
This fixes the advanced time evolution of the gauge parameter $\epsilon$ to
\begin{subequations}
\begin{align}
 \partial_u \epsilon_M^{\pm} & =  \pm \epsilon_L^{\pm} \mu_M^0 \mp \epsilon_L^0 \mu_M^{\pm}  \pm \epsilon_M^{\pm} \mu_L^0 \mp \epsilon_M^0 \mu_L^{\pm} \\
\frac12 \partial_u \epsilon_M^{0} & = \epsilon_L^+ \mu_M^{-} - \epsilon_L^{-} \mu_M^+ + \epsilon_M^+ \mu_L^{-} - \epsilon_M^{-} \mu_L^+  \\
 \partial_u \epsilon_L^{\pm} & =  \pm \epsilon_L^{\pm} \mu_L^0 \mp \epsilon_L^0 \mu_L^{\pm} \\
\frac12 \partial_u \epsilon_L^{0} & = \epsilon_L^+ \mu_L^{-} - \epsilon_L^{-} \mu_L^+.
\end{align}
\end{subequations}
The variation of the canonical boundary charge in this case is then given by
\begin{equation}
\delta Q[\epsilon] = \frac{k}{\pi} \oint \extd \varphi \left( \epsilon_M^+ \delta \mathcal{M}^- + \epsilon_M^0 \delta\mathcal{M}^0 + \epsilon_M^- \delta\mathcal{M}^+ +\epsilon_L^+ \delta\mathcal{L}^- + \epsilon_L^0 \delta\mathcal{L}^0 + \epsilon_L^- \delta\mathcal{L}^+ \right) \,.
\end{equation}
Assuming that the gauge parameter $\epsilon$ does not depend on any of the functions $\mathcal{L}^a, \mathcal{M}^a$ one can trivially integrate the charges in field space to obtain
\begin{equation}
Q[\epsilon] = \frac{k}{\pi} \oint \extd \varphi  \left( \epsilon_M^+ \mathcal{M}^- + \epsilon_M^0 \mathcal{M}^0 + \epsilon_M^- \mathcal{M}^+ +\epsilon_L^+ \mathcal{L}^- + \epsilon_L^0 \mathcal{L}^0 + \epsilon_L^- \mathcal{L}^+ \right).
\end{equation}
The algebra of asymptotic symmetries can then be obtained again in the same way as in the previous sections. In terms of the Fourier modes
\begin{equation}
\mathcal{L}_n^a = \frac{k}{\pi} \oint \extd\varphi e^{-in\varphi} \mathcal{L}^a \qquad 
\mathcal{M}_n^a = \frac{k}{\pi} \oint \extd\varphi e^{-in\varphi} \mathcal{M}^a
\end{equation}
one obtains the Dirac brackets algebra
\begin{subequations}
\begin{align}
\{\mathcal{L}_n^a , \mathcal{L}_m^b \} & = (a-b)\mathcal{L}_{n+m}^{a+b} \\
\{\mathcal{L}_n^a , \mathcal{M}_m^b \} & = (a-b)\mathcal{M}_{n+m}^{a+b} -  i n k \kappa_{ab}\delta_{n+m,0} \\
\{\mathcal{M}_n^a , \mathcal{M}_m^b \} & = 0\,.
\end{align}
\end{subequations}
Taking $M_n^a = i \mathcal{M}_n^a, L_n^a = i \mathcal{L}_n^a$ and replacing the Dirac brackets by commutators $i\{ \, , \} = [\, , ]$ yields the commutator algebra
\begin{subequations}
\begin{align}
[L_n^a , L_m^b] & = (a-b)L_{n+m}^{a+b} \\
[L_n^a , M_m^b] & = (a-b)M_{n+m}^{a+b} -  n k \kappa_{ab}\delta_{n+m,0} \\
[M_n^a , M_m^b] & = 0\,.
\end{align}
\end{subequations}
This is the affine $\mathfrak{isl}(2)_k$ algebra. What we see here is that, indeed, in three-dimensional pure Einstein-Hilbert gravity all the relevant physical degrees of freedom are encoded in the boundary. Even though we started with the exact same bulk theory, just a different choice of boundary conditions yielded a completely different asymptotic symmetry algebra as before. As an addendum it should also be noted that even though the boundary conditions that lead to $\mathfrak{bms}_3$ as the asymptotic symmetry algebra are a subset of the boundary conditions presented here, the resulting affine $\mathfrak{isl}(2)_k$ algebra does not contain $\mathfrak{bms}_3$ as a subalgebra. This has to do with the fact that this whole procedure of specifying boundary conditions and determining the asymptotic symmetries of the canonical boundary charges is just a Hamiltonian reductions of $\mathfrak{isl}(2,\mathbb{R})$ in disguise and as such the relation between subsets of a given set of boundary conditions and subsets of the resulting asymptotic symmetry algebras is not really in one-to-one correspondence.

\section{Soft Hair in 3D}

This section of the lecture notes is devoted to a certain set of boundary conditions in three-dimensional Einstein gravity that can be interpreted as a black hole carrying soft hair excitations\footnote{Usually excitations that carry zero energy are called soft excitations. Of particular interest for the purpose of these lecture notes will be excitations that are interpreted as supertranslation hair of black holes.}. We will first review near-horizon boundary conditions that have been first described in the context of three-dimensional Einstein gravity\footnote{For similar boundary conditions in the context of Chern-Simons-like theories of gravity or Generalized Minimal Massive Gravity see e.g. \cite{Setare:2016vhy,Setare:2016jba}.} \cite{Afshar:2016wfy} as well as the resulting near-horizon symmetry algebra. Following up on this we also include a brief discussion as to why studying soft excitations in the context of the black hole information paradox is an interesting thing to do.

\subsection{Near-Horizon Boundary Conditions}

Up until now all the considerations in these lecture notes were focused on \emph{asymptotic} symmetries. That is, imposing boundary conditions at asymptotic infinity of a given spacetime. What we will now review in the following can be seen as kind of the the opposite programme. Non-extremal black holes can be universally approximated by a product of two-dimensional Rindler space \cite{Rindler:1966zz} with a compact Euclidean manifold. For the special case of three-dimensional gravity with cosmological constant $\Lambda=-\frac{1}{\ell^2}$ this means that the metric around a non-extremal black hole can be describen in terms of ingoing Eddington-Finkelstein coordinates as
    \begin{equation}\label{eq:SoftHairNHBCsMetric}
    \extd s^2=-2a\ell\rho f\extd\nu^2+2\ell\extd\nu\extd\rho-2\frac{\omega}{a}\extd\varphi\extd\rho+4\omega\rho f\extd\nu\extd\varphi+\left[\gamma^2+\frac{2\rho}{a\ell}f\left(\gamma^2-\omega^2\right)\right]\extd\varphi^2,
    \end{equation}
where $\ell\rho=r$ and $f:=1+\frac{\rho}{2a\ell}$. The horizon is located at $r=0$, $\nu$ is the advanced time and the angular coordinate $\varphi$ is assumed to be $2\pi$-periodic, i.e. $\varphi\sim\varphi+2\pi$. The parameter $a$ is the Rindler acceleration and is assumed to be constant in contrast to the functions $\omega$ and $\gamma$ that are assumed to be arbitrary functions of $\varphi$.\\
This line element is a solution of Einsteins equations in three dimensions with constant negative curvature and describes the near-horizon physics of black holes that are in general not spherically symmetric. Thus such solutions are also called black flowers \cite{Barnich:2015dvt}. For the case of constant $\omega$ and $\gamma$ these solutions reduce to the well known BTZ black hole \cite{Banados:1992wn,Banados:1992gq}.\\
A remarkably simple way of writing down near-horizon boundary conditions that obey \eqref{eq:SoftHairNHBCsMetric} was described in \cite{Afshar:2016wfy} using the Chern-Simons formulation. In order to describe AdS$_3$ gravity in three dimensions the Chern-Simons coupling $k$ has to be related to Newton's constant $G$ as alreay described in \eqref{eq:ChernSimonsLevelNewtonConstant}. In addition the Chern-Simons connection $\mathcal{A}$ has to take values in $\mathfrak{sl}(2,\mathbb{R})\oplus\mathfrak{sl}(2,\mathbb{R})$ and thus can be split into two connections $A^\pm$ that each take values in a single copy of $\mathfrak{sl}(2,\mathbb{R})$. The three generators of $\mathfrak{sl}(2,\mathbb{R})$ are chosen in such a way that they obey
    \begin{equation}
        [\Lt_n,\Lt_m]=(n-m)\Lt_{n+m},
    \end{equation}
for $n,m=\pm1,0$. In this basis the invariant bilinear form is then given by
	\begin{subequations}\label{eq:SLInvBilForm}
	\begin{align}
		\langle \Lt_n\Lt_m\rangle & =-\left(
			\begin{array}{c|ccc}
				  &\Lt_1&\Lt_0&\Lt_{-1}\\
				\hline
				\Lt_1&0&0&1\\
				\Lt_0&0&-\frac{1}{2}&0\\
				\Lt_{-1}&1&0&0
			\end{array}\right).
	\end{align}		
	\end{subequations}
A metric formualtion can be obtained from the Chern-Simons connections via
    \begin{equation}\label{eq:ConnectionToMetric}
        g_{\mu\nu}=\frac{\ell^2}{2}\left\langle(A^+_\mu-A^-_\mu)(A^+_\nu-A^-_\nu)\right\rangle.
    \end{equation}
Looking at the metric \eqref{eq:SoftHairNHBCsMetric} and \eqref{eq:ConnectionToMetric} a natural choice of boundary conditions is given by
    \begin{equation}
        A^\pm=b^{-1}_\pm(\extd+\mathfrak{a}^\pm)b_\pm,
    \end{equation}
where
    \begin{equation}
        b_\pm=e^{\pm\frac{1}{\ell\zeta^\pm}L_1}e^{\pm\frac{\rho}{2}L_{-1}},
    \end{equation}
and
    \begin{equation}\label{eq:SoftHairBCs}
        \mathfrak{a}^\pm=\left(\pm\mathcal{J}^\pm\extd\varphi+\zeta^\pm\extd\nu\right)L_0,
    \end{equation}
with $\ell\mathcal{J}^\pm:=\gamma\pm\omega$. In general the state dependent functions $\mathcal{J}^\pm$ as well as the (fixed\footnote{Fixed in that context means that the variation is zero, i.e. $\delta\zeta^\pm=0$.}) chemical potentials $\zeta^\pm$ are a priori arbitray functions of $\varphi$ and $\nu$. However, imposing the equations of motion $F=0$ one obtains the following relations:
    \begin{equation}
    \partial_\nu\mathcal{J}^\pm=\pm\partial_\varphi\zeta^\pm.
    \end{equation}
One can assume for simplicity that the chemical potentials are constant. Then the previous relations fix the state dependent functions $\mathcal{J}^\pm$ to be arbitrary functions of only $\mathcal{\varphi}$. For the specific case of $\zeta^\pm=-a$ one recovers exactly \eqref{eq:SoftHairNHBCsMetric}.

\subsection{Symmetry Algebra}

The rest of the procedure to determine the near-horizon symmetry algebra is exactly the same as described previously in these lecture notes. The first step is to determine the boundary condition preserving gauge transformations i.e. all gauge transformations that satisfy     \begin{equation}
        \delta_{\epsilon^\pm}\mathfrak{a}^\pm=\extd\epsilon^\pm+[\mathfrak{a}^\pm,\epsilon^\pm]=\mathcal{O}(\delta\mathfrak{a}^\pm),
    \end{equation}
for some gauge parameters $\epsilon^\pm$. Making the ansatz $\epsilon^\pm=\epsilon^\pm_n \Lt_n$ one finds that the variation of the canonical boundary charges is given by
    \begin{equation}
        \delta Q[\epsilon^\pm]=\pm\frac{k}{4\pi}\int\extd\varphi\eta^\pm\delta\mathcal{J}^\pm,
    \end{equation}
where $\epsilon^\pm_0=\eta^\pm$. The state dependent functions $\mathcal{J}^\pm$ transform under these gauge transformations as
    \begin{equation}\label{eq:HairyVariations}
        \delta_{\eta^\pm}\mathcal{J}^\pm=\pm\partial_\varphi\eta^\pm.
    \end{equation}
The variation of the boundary charges can be trivially functionally integrated to yield
    \begin{equation}\label{eq:HairyCharges}
        Q[\epsilon^\pm]=\pm\frac{k}{4\pi}\int\extd\varphi\eta^\pm\mathcal{J}^\pm.
    \end{equation}
Using the same arguments that were already presented in Section~\ref{sec:FSASA} one can directly determine the Dirac brackets of the state dependent functions from \eqref{eq:HairyVariations} and \eqref{eq:HairyCharges}. Using the Fourier mode expansion
    \begin{equation}
        J^\pm_n=\frac{k}{4\pi}\int\extd\varphi\mathcal{J}^\pm(\varphi)e^{in\varphi},
    \end{equation}
one obtains the following near-horizon symmetry algebra:
    \begin{equation}
        [J^\pm_n,J^\pm_m]=\pm\frac{k}{2}n\delta_{n+m,0},
    \end{equation}
where in addition $[J^+_n,J^-_m]=0$. This is a remarkably simple symmetry algebra as it consists of two affine $\hat{\mathfrak{u}}(1)$ symmetry algebras with level $\pm\frac{k}{2}$.\\
After having found the near-horizon symmetry algebra corresponding to the boundary conditions \eqref{eq:SoftHairBCs} the next question to answer is why it is justified to interpret these boundary conditions as describing soft hair excitations. The key to answering this question lies in the definition of soft hair being zero-energy excitations of the vacuum. For that purpose we first have to determine the Hamiltonian governing time evolution in our setup.\\
Time evolution in gravitational systems is usually covered by a timelike Killing vector. In our concrete setup this would mean a Killing vector along the advanced time $\nu$. Recalling that on-shell the gauge parameters in the Chern-Simons formulation and the asymptotic Killing vectors are related via $\epsilon^+-\epsilon^-=\xi^\mu(A^+_\mu-A^-_\mu)$ one finds that the variation of the canonical charge associated to the Killing vector $\xi^\nu$ is given by
    \begin{equation}
        \delta \mathcal{H}=\delta Q[\epsilon^+]-\delta Q[\epsilon^-]=\frac{k}{4\pi}\int\extd\varphi\left\langle\xi^\nu\left(A^+_\nu \delta A^+_\varphi-A^-_\nu \delta A^-_\varphi\right)\right\rangle.
    \end{equation}
For the particular choice of $\zeta^\pm=-a$ one obtains as the Hamiltonian $H=-a(J^+_0+J^-_0)$.\\
The crucial thing to note here is that the Hamiltonian is in the center of the near-horizon symmetry algebra. Thus it also commutes with all other generators. Assume now that one can build any quantum state by exciting the vacuum by acting arbitrarily with the generators $J^\pm_n$ with $n<0$ on a vacuum state $|0\rangle$. Since $\mathcal{H}$ commutes with all $J^\pm_n$ generators it also follows that any excited state that is obtained via acting with the near-horizon symmetry generators on the vacuum has exactly the same energy as the vacuum. Thus all excitations in this module are zero energy excitations of the vacuum and it is thus sensible to call them soft hair in the sense of \cite{Hawking:2016msc}.

\subsection{Why are Soft Excitations Interesting?}

Even thought we have explained what soft excitations are in the previous section we did neither point out their physical relevance nor the reason why the term ``soft hair'' has gained so much attention during the past one and a half years. The purpose of this section is to deal with this shortcoming and provide a brief overview of the possible consequences of black holes having soft excitations.\\
The main reason why the existence of soft hair is an exciting prospect within the context of (quantum) gravity is that it may provide a possible solution to the black hole information paradox. This paradox deals with the fundamental question what happens with the information contained within the black hole during its evaporation process? In 1975 Stephen Hawking argued that the information will be lost in the course of the black hole evaporating \cite{Hawking:1974sw,Hawking:1976ra}. Since then there has been a lot of research going on in trying to solve this paradox. In 2016 Hawking, Perry and Strominger pointed out in \cite{Hawking:2016msc} that maybe two of the underlying assumptions leading to the conclusion that information is lost were incorrect. The first one being that the vacuum state in quantum gravity is unique and the second one being that black holes do not have any hair i.e. black holes are completely determined in terms of their mass, angular momentum and electric charge.\\
What lead to the suspicion that there might be a loophole to the original arguments by Hawking was an observation made in \cite{Strominger:2013jfa}. This observation was that there was an infinite amount of conservation laws governing the scattering of gravitons. These infinite conservation laws are given by the supertranslation generators that are part of the $\mathfrak{bms}$ algebra that governs the asymptotic symmetries of asymptotically flat spacetimes. The interesting thing now is that acting with supertranslations on a given state excites that state, however, with zero energy difference. So the new state has the same energy as before, but is physically distinct from the previous state. This is again closely related to the fact that the asymptotic charges associated of diffeomorphisms that are associated with supertranslations are non-zero and thus are what we called \emph{improper} gauge transformations. What Hawing-Perry-Strominger argued in \cite{Hawking:2016msc} as well as \cite{Hawking:2016sgy} was that by acting on a black hole horizon with supertranslations one basically adds photons with zero energy to the black hole horizon that can be considered as soft hair. If a particle now falls into the black hole this soft hair can be excited by that process. Since these charges have to be conserved because of supertranslation invariance this in turn also means that the information that entered the black hole should not be completely lost. Thus soft excitations may provide a new angle on finding a possible solution to the black hole information paradox\footnote{See also e.g. \cite{Strominger:2017aeh}. It should also be noted that some authors also argue that soft hair might not solve the black hole information paradox. See e.g. \cite{Bousso:2017dny,Bousso:2017rsx} and references therein.}.

\acknowledgments{We would like to thank the organizing committee of the XIII Modave School in Mathematical Physics for inviting us to give these series of lecture. The research of MR is supported by the ERC Starting Grant 335146 "HoloBHC". CZ thanks G. Comp\`ere, V. Lekeu, A. Marzolla. She is a research fellow of "Fonds pour la Formation \`a la Recherche dans l'Industrie et dans l'Agriculture"-FRIA Belgium. This work is partially supported by FNRS-Belgium (convention IISN 4.4503.15).}

\bibliographystyle{JHEP}
\bibliography{Bibliography}

\end{document}